\documentclass[traditabstract]{aachanged}
\usepackage[english]{babel}
\usepackage{amsmath}
\usepackage{amssymb}
\usepackage{euscript}
\usepackage[dvips]{graphicx}
\usepackage[T1]{fontenc}
\usepackage{aecompl} 
\usepackage{cyr}
\usepackage{epsfig}
\usepackage{multirow} 
\begin{document}

\title{Estimation of Physical Conditions in the Cold Phase of the Interstellar
Medium in the sub-DLA System at $z = 2.06$ in the Spectrum
of the Quasar J\,2123$-$0050}
\titlerunning{Estimation of Physical Conditions}
\author{V.V. Klimenko$^{1,2}$\thanks{E-mail: slava.klimenko@gmail.com}, S.A. Balashev$^{1,2}$, A.V. Ivanchik$^{1,2}$, D.A. Varshalovich$^{1,2}$}
\authorrunning{Klimenko et al.}
\date{Received 21 October, 2015}

\institute{\it{$^{1}$Ioffe Physicotechnical Institute, Russian Academy of Sciences, ul. Politekhnicheskaya 26,
St. Petersburg, 194021 Russia} \\
\it{$^{2}$St. Peter the Great St. Petersburg Polytechnic University, Polytechnicheskaya ul. 29, St. Petersburg, 195251 Russia
Received October 21, 2015}}



\abstract{An independent analysis of the molecular hydrogen absorption system at redshift $z_{\rm abs} = 2.059$ in the spectrum of the quasar J\,2123$-$0050 is presented. The H$_2$ system consists of two components (A and B) with column densities $\log N^{\rm A}_{\rm H_2} = 17.94\pm0.01$ and $\log N^{\rm B}_{\rm H_2} = 15.16\pm0.02$. The spectrum exhibits the lines of HD molecules ($\log N^{\rm A}_{\rm HD} = 13.87\pm0.06$) and the neutral species C\,{\sc i} and Cl\,{\sc i} associated with the H$_2$ absorption system. For the molecular hydrogen lines near the quasar's Ly$\beta$ and O\,{\sc vi} emission lines, we detect a nonzero residual flux, $\sim3\,\%$ of the total flux, caused by the effect of partial coverage of the quasar's broad-line region by an H$_2$ cloud. Due to the smallness of the residual flux, the effect does not affect the H$_2$ column density being determined but increases the statistics of observations of the partial coverage
effect to four cases. The uniqueness of the system being investigated is manifested in a high abundance of the neutral species H$_2$ and C\,{\sc i} at the lowest H\,{\sc i} column density, $\log N_{\rm H\,I} = 19.18\pm0.15$, among the high redshift systems. The H$_2$ and C\,{\sc i} column densities in the system being investigated turn out to be higher than those in similar systems in our Galaxy and theMagellanic Clouds by two or three orders ofmagnitude. The $N_{\rm HD}/2N_{\rm H_2}$ ratio for component A has turned out to be also unusually high, $(4.26\pm0.60)\times10^{-5}$, which exceeds the deuterium abundance (D/H) for high-redshift systems by a factor of 1.5. Using the H\,{\sc i}, H$_2$, HD, and C\,{\sc i} column densities and the populations of excited H$_2$ and C\,{\sc i} levels, we have investigated the physical conditions in components A and B. Component A represents the optically thick case; the gas has a low number density ($n\approx30\,cm^{-3}$) and a temperature $T\sim140$\,K. In component B, the mediumis optically thin with $n<100\,{\rm cm}^{-3}$ and $T\ge100$\,K. The ultraviolet (UV) background intensity in the clouds exceeds the mean intensity in our Galaxy by almost an order ofmagnitude. A high gas ionization fraction, $n_{\rm H^+}/n_{\rm H}\sim 10^{-2}$, which can be the result of partial shielding of the systemfrom hard UV radiation, is needed to describe the high HD and C\,{\sc i} column densities. Using our simulations with the PDR\,Meudon code, we can reconstruct the observed column densities of the species within the model with a constant density ($n_{\rm H}\sim40\,{\rm cm}^{-3}$). A high H$_2$ formation rate (higher than the mean Galactic value by a factor of $10-40$) and high gas ionization fraction and UV background intensity are needed in this case.}

\keywords{
galaxies, interstellar medium, absorption systems, quasar J\,2123$-$0050}

\maketitle

\section{Introduction}
\label{introduction}
\noindent
One way to study the physical conditions and
chemical composition of the interstellar and intergalactic
media at high redshifts is to investigate the
absorption systems in the spectra of bright extragalactic
sources, such as quasars or gamma-ray
bursts. Most of the absorption lines in the spectra
of such objects are associated with the Lyman alpha
(Ly$\alpha$) lines of neutral hydrogen falling within the
optical range due to the cosmological redshift $z$. The
absorption systems of the Ly$\alpha$ forest are believed to
relate to the clouds of almost fully ionized hydrogen
located in the intergalactic medium on the line of
sight between the quasar and the observer. Neutral
hydrogen absorption systems with a very high H\,{\sc i}
column density ($\log N_{\rm H\,I} > 19$, here and below, the
column densities are measured in cm$^{-2}$) are detected
approximately in 10\,\% of the quasar spectra. If
the column density $\log N_{\rm H\,I} > 20.3$, then such an
absorption system is called a DLA system (Damped
Lyman Alpha system); if $19.0 < \log N_{\rm H\,I}< 20.3$, then
the system belongs to the class of sub-DLA systems.
The DLA systems are believed to be the main
reservoirs of neutral hydrogen in the early Universe
\cite{Prochaska2009, Noterdaeme2012}.
In the spectra of quasars, the sub-DLA and DLA systems
are identified owing to the broad Lyб absorption
line with characteristic Lorentz wings. Another
peculiarity of such systems is the presence of many
absorption lines of heavy elements\footnote{Elements heavier than helium are called metals.} in the spectrum
at the same redshift. The DLA systems appear to
be the disks or halos of protogalaxies with a radius
of several 10 kpc \cite{Krogager2012}. Due
to the high column density $N_{\rm H\,I}$, the matter in DLA
systems is shielded from ionizing ultraviolet (UV)
radiation with an energy exceeding the ionization
energy of the hydrogen, $E > 13.6$\,eV. But, at
the same time, it is believed that the medium in
sub-DLA systems can be partially shielded due to
the lower HI column density, and gas regions with
different ionization fractions can simultaneously fall
on the line of sight. When the elemental abundances
in sub-DLA systems are analyzed, the 
ionization corrections should be applied (see, e.g.,
\cite{Meiring2007, Milutinovic2010}).

The detection of the H$_2$, HD, and CO molecular
absorption systems (see \cite{Levshakov1985, Varshalovich2001, Srianand2008, Ivanchik2015}), which are known
from observations in our Galaxy and Local Group galaxies (see, e.g., \cite{Jenkins2001, Snow2008, Rachford2008}) relate to dense
cold clouds ($n\sim10-500$\,cm$^{-3}$, $T\sim40-200$\,K) in the interstellar medium, suggests that the matter in
DLA and sub-DLA systems belongs to the interstellar medium and not to the intergalactic one.

\begin{figure*}
\begin{center}
        \includegraphics[width=1.0\textwidth]{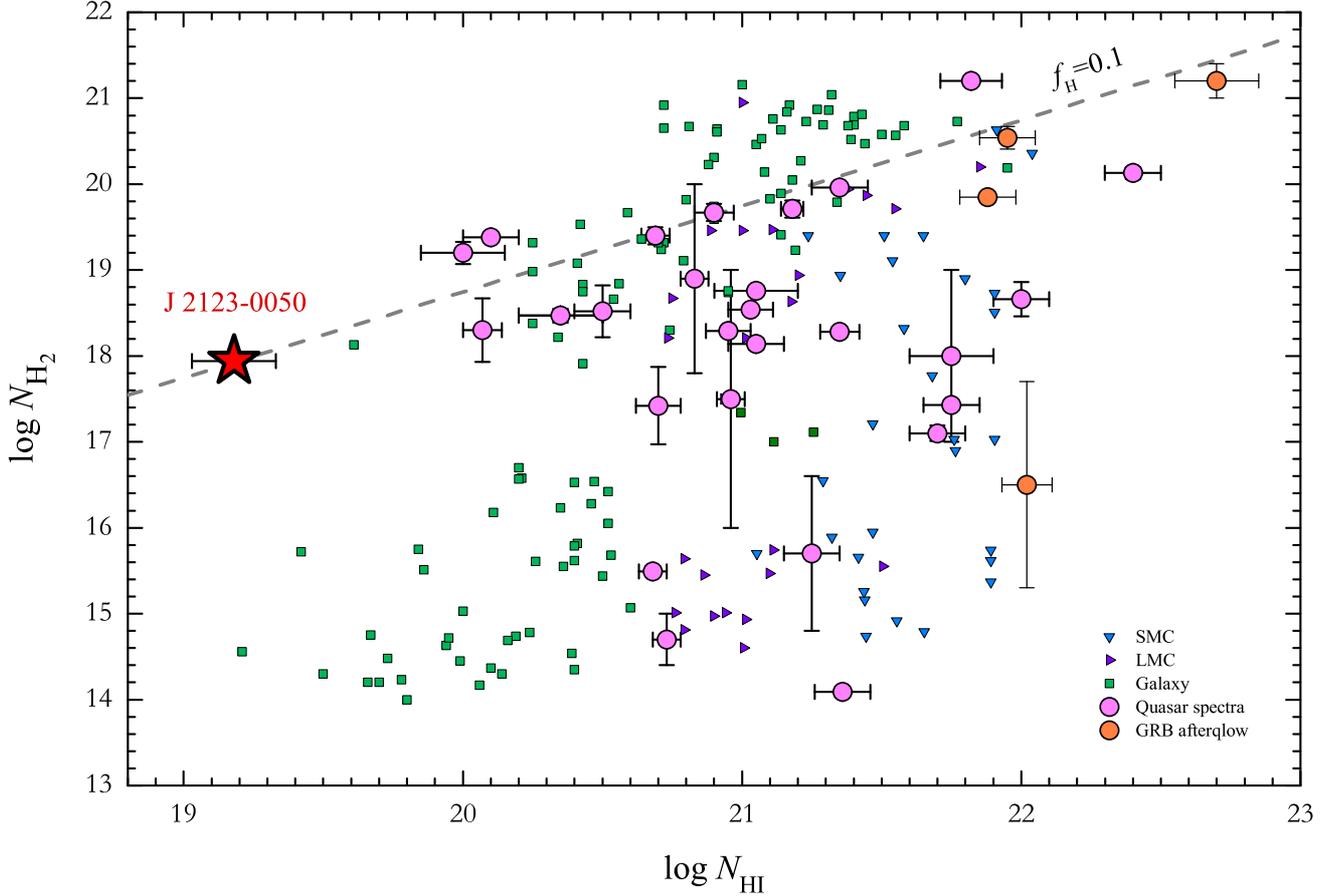}
        \caption{Measured H$_2$ and H\,{\sc i} column densities for the high-redshift DLA and sub-DLA systems identified in the spectra of
quasars (filled dark circles -- see, e.g., \cite{Ivanchik2015, Noterdaeme2015}) and the spectra of GRB afterglows (filled
light circles -- see \cite{Prochaska2009, Kruhler2013, DElia2014, Friis2015}) as well as for the systems
in our Galaxy and the Magellanic Clouds (squares, diamonds, and triangles -- \cite{Rachford2008, Welty2012}). The system being investigated has the lowest HI column density ($\log N_{\rm H\,I} = 19.18\pm0.15$) among the high-redshift systems with $\log N_{\rm H_2}
\sim 18$.}
        \label{H2vsHI} 
\end{center}
\end{figure*}

The H$_2$ absorption systems are observed in less than 10\,\% of the spectra containing DLA and sub-DLA systems \cite{Noterdaeme2008, Balashev2014, Jorgenson2014}. Spectra with a high signal-to-noise ratio ($S/N>10$) and a high spectral resolution ($R\sim20\,000-110\,000$), which is a limiting problem for the largest optical telescopes at present, are needed for their investigation. Thirty high-redshift systems have been detected (see, e.g., \cite{Ivanchik2015, Noterdaeme2015}) since the first identification of H$_2$ molecular lines in the spectra of quasars \cite{Levshakov1985}, despite the fact that more than 12\,000 DLA systems are
already known \cite{Noterdaeme2012}. We have developed a technique of searching for H$_2$ candidates in the medium-resolution spectra of the Sloan Digital Sky Survey (SDSS, \cite{York2000}), which has revealed more than 50 H$_2$ candidates to date \cite{Balashev2014}. Moreover, the first observations of eight candidates at the Very Large Telescope (VLT) led to their confirmation (Balashev et al., in preparation). The main objective of studying high-redshift molecular hydrogen clouds is to determine the physical conditions in the cold phase of the interstellar medium. This is important for understanding the star formation and evolution of galaxies at high redshifts and for understanding the nature of DLA systems and their connection with galaxies.

In this paper, we present an independent analysis of the H$_2$/HD absorption system at redshift $z_{\rm abs} = 2.059$ in the VLT spectrum of the quasar J\,2123$-$0050. Previously, this system has already been investigated in the quasar spectrum taken at the Keck telescope. However, the results of two papers \cite{Malec2010, Tumlinson2010} disagree both in the number of components of the investigated absorption system and in the estimated H$_2$ column density. A high signal-to-noise ratio in continuum in the VLT spectrum, $S/N\sim 20-120$, compared to $S/N\sim10-40$ in the Keck spectrum allowed one to study in more detail the structure of this system and to obtain more accurate column densities of H$_2$ and HD molecules at various rotational levels. 

An unusual feature of the H$_2$ system in the spectrum of J\,2123$-$0050 is a high H$_2$ column density
for one of the components of the absorption system, $\log N^{\rm A}_{\rm H_2} = 17.94\pm0.01$, despite the fact that the atomic hydrogen column density $\log N_{\rm H\,I} = 19.18\pm0.15$ \cite{Milutinovic2010} is lowest among the DLA and sub-DLA systems in which the H$_2$ systems were detected (see Fig.\,\ref{H2vsHI}). The measured H$_2$ column density in the spectrum of J\,2123$-$0050 is almost three orders of magnitude higher than that for the H$_2$ clouds located in our Galaxy and having similar column densities $N_{\rm H\,I}$. One would think that such a difference can be explained by a relatively low UV background intensity. However, it has been found that the UV background intensity in the sub-DLA system of J\,2123$-$0050 to be considerably higher than the mean Galactic value, \cite{Milutinovic2010}, \cite{Som2013}.

Apart from the H$_2$ molecules, the lines of HD molecules are detected in the system J\,2123$-$0050 \cite{Tumlinson2010}. The $N_{\rm HD}/2N_{\rm H_2}$ ratio in this system turned out to be a factor of 1.5 higher than the primordial deuterium abundance and a factor
of 3 or 4 higher than the values measured in high redshift H$_2$/HD systems (see \cite{Ivanchik2015}). These and other facts are of additional interest for the studies of physical conditions in this H$_2$ system.

\section{OBSERVATIONAL DATA}

The quasar J\,2123$-$0050 (with redshift $z_{\rm em} = 2.261$
and apparent magnitude $m_{\rm V} = 16.6$) was identified
in the SDSS. High-resolution spectra of this quasar
were taken independently at the two largest optical
telescopes, Keck (using the High Resolution Echelle
Spectrograph, HIRES) and VLT (using the Ultraviolet
and Visible Echelle Spectrograph, UVES).
The quasar was observed at the Keck telescope in
the wavelength range $3071-5869$\,\AA\, in 2006 under
the U080Hb program (Prochaska). The data are
in free access from the Keck archive\footnote{https://koa.ipac.caltech.edu/cgi-bin/KOA/nph-KOAlogin} . The slit
size was chosen to be 0.4'', the CCD pixels were
binned $2\times1$. These settings allowed a spectrum
with a resolution $R\sim100\,000$ to be achieved. We
reduced and added the exposures using the MAKEE
software package\footnote{http://www.astro.caltech.edu/ tb/makee} . The total exposure time was
$\sim6.1$\,h. J\,2123$-$0050 was observed at the VLT in 2008 under the 81.A-0242 program (Ubachs). The
data are in free access from the European Souther Observatory (ESO) archive4 . Different settings were
used for the blue and red arms of the UVES spectrograph. For the blue arm, the slit size was chosen to be 0.8'', and the CCD pixels were binned $2\times2$. For the red arm, the slit size was chosen to be 0.7'', and the CCD pixels were binned $1\times1$.
These settings allows a spectrum with a resolution of 49\,620 for the blue part and 56\,990 for the red one to be achieved. We reduced the individual exposures using the UVES Common Pipeline Library (CPL) data reduction pipeline release 4.9.5 software package\footnote{ftp://ftp.eso.org/pub/dfs/pipelines/uves/uves-pipelinemanual-22.8.pdf}. The total exposure time was $\sim11.3$\,h.

\section{THE SUB-DLA ABSORPTION SYSTEM AT $z = 2.06$}

The sub-DLA system at redshift $z_{\rm abs} = 2.06$ in the
spectrum of the quasar J\,2123$-$0050 was first investigated by \cite{Milutinovic2010} and
then by \cite{Som2013}. The sub-DLA system consists of at least two components with column
densities $\log N_{\rm H\,I}(z = 2.05684) = 18.40\pm0.30$ and $\log N_{\rm H\,I}(z = 2.05930) = 19.18\pm0.15$ \cite{Milutinovic2010}. The lines of H$_2$, HD, C\,{\sc i}, and Cl\,{\sc i} molecules, which are known to be indicators of the cold phase of the interstellar medium (see, e.g., \cite{Jura1974, Noterdaeme2007, Balashev2015}), are detected in the component at $z =2.05930$.

\subsection{The Ionization Structure}

In it well known that in contrast to DLA systems,
where the medium is thought to be predominantly
neutral, much of the hydrogen (up to 90\%)
in sub DLA systems is in an ionized state, which
should be taken into account when analyzing such
systems. The sub-DLA systems are believed to have
a layered (multiphase) structure. The hydrogen ionization
fractions in the neutral and ionized phases
are close to 0 and 1, respectively, i.e., the matter is
separated into predominantly fully neutral and fully
ionized phases and is not a mixture with some ionization
fraction.

The metals in sub-DLA systems occupy a spatially
larger region than does the neutral hydrogen;
therefore, the structures of sub-DLA systems (H\,{\sc i}
+ H\,{\sc ii}) manifest themselves in the absorption lines
of metals with different ionization states, such as
O, S, Si, Al, Zn, etc. In a predominantly neutral
medium, the metals are in the ground state, at the first
ionization level with an energy above 13.6 eV, because
the UV photons capable of ionizing such ions are
absorbed by neutral hydrogen. In the region of almost
fully ionized hydrogen in sub-DLA systems, the
relationship between the number densities of metals
at various ionization levels is established in accordance
with the balance between the recombination
and ionization rates of the species. In the spectrum,
such an ionization structure manifests itself in the
simultaneous presence of absorption lines of metals
with various ionization states in the same velocity
components.

\begin{figure*}
\begin{center}
        \includegraphics[width=1.0\textwidth]{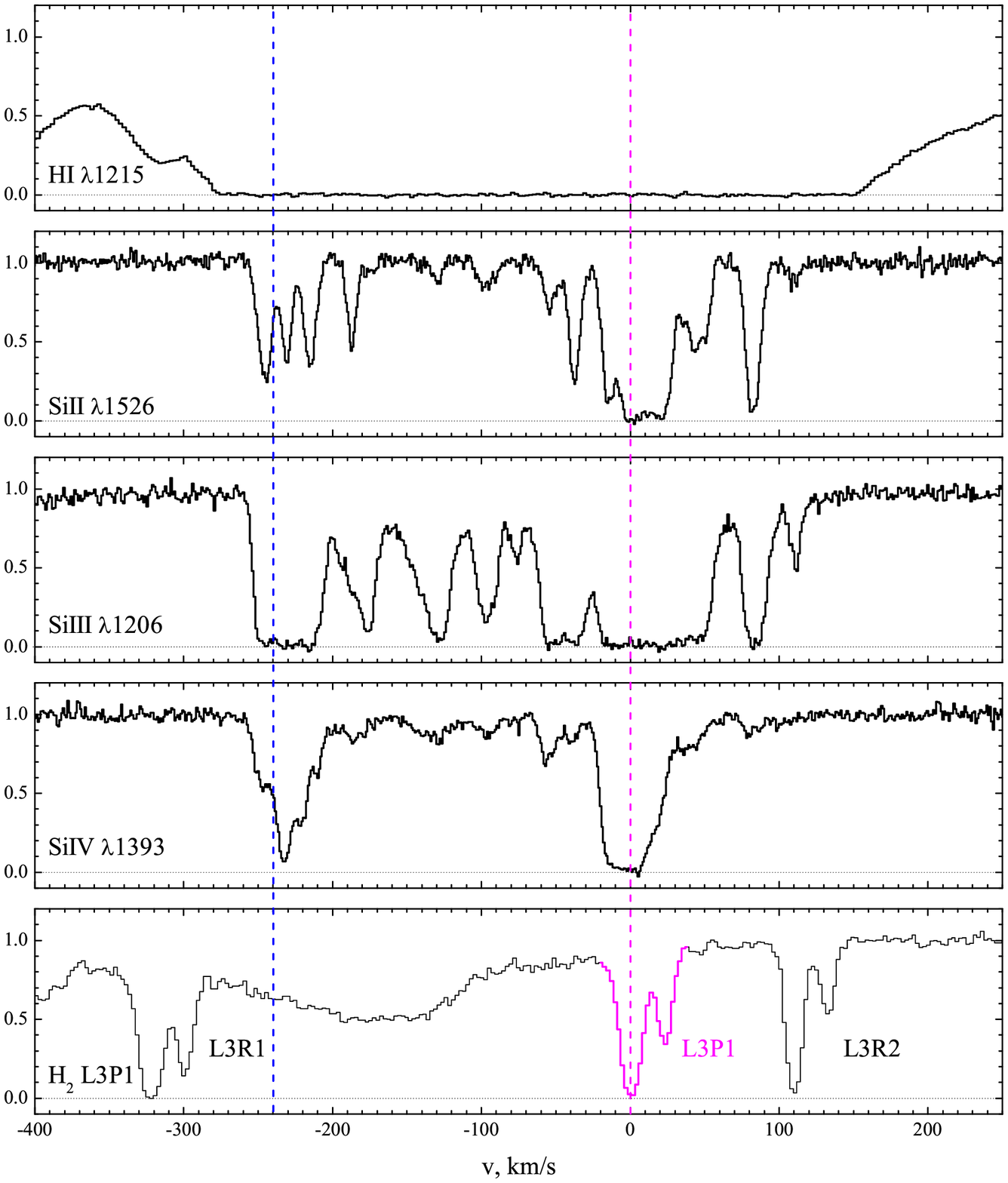}
        \caption{H\,{\sc i} (Ly$\alpha$), Si\,{\sc ii}, Si\,{\sc iii}, Si\,{\sc iv}, and H$_2$ absorption line profiles. The dashed vertical lines indicate the positions of the two components of the sub-DLA system at $z = 2.05684$ and $2.05930$. The lower panel shows the H$_2$ L3P1 line profile. The two components of the H$_2$ absorption system corresponding to the component of the sub-DLA system at $z = 2.05930$ are highlighted by the thick line. The H$_2$ lines corresponding to the L3R1 and L3R2 transitions are also specified.}
        \label{DLAstr} 
\end{center}
\end{figure*}

In the sub-DLA system in the spectrum of
J\,2123$-$0050, for example, Si is represented at
least in three ionization states, Si\,{\sc ii}, Si\,{\sc iii}, and Si\,{\sc iv},
whose absorption structure is shown in Fig.\,\ref{DLAstr}. The
system has up to 20 individual components detected
simultaneously in Si (and Al) lines in various ionization
states. Given the ionization corrections, the
metallicity changes from ${\rm [Si/H] = +0.00\pm0.02}$\footnote{${\rm [X/H]} = \log(N({\rm X})/N({\rm H})) - \log(N({\rm X})/N({\rm H}))_{\odot}$ is the difference between the logarithms of the elemental abundance in the system being investigated and with respect to the
abundance measured in the Solar system.} to ${\rm [Si/H] = -0.71\pm0.15}$ \cite{Milutinovic2010}.

\section{THE MOLECULAR HYDROGEN SYSTEM}

The H$_2$ system in the spectrum of the quasar J\,2123$-$0050 has been investigated in several studies. The lines of H$_2$ and HD molecules were studied independently by \cite{Malec2010} in the Keck spectrum and by \cite{vanWeerdenburg2011} in the VLT spectrum to place constraints on the possible time variation of the proton-to-electron mass ratio. The HD/2H$_2$ ratio of the molecular cloud was
investigated by \cite{Tumlinson2010} in the Keck spectrum (for a comparison of the results of these analyses, see Table\,\ref{table_1}). We performed an independent analysis of this absorption system in the VLT spectrum. The parameters of the absorption system were determined by comparing the observed spectrum with the synthetic one. To determine the best-fit
parameters, we used the Markov Chain Monte Carlo (MCMC) technique.

\begin{table*}
\begin{center}
\caption{Comparison of the measured H$_2$ and HD column densities in component A of the H$_2$ absorption system
at $z_{\rm abs} = 2.059$ in the spectrum of J\,2123$-$0050.} 
\label{table_1}
\begin{tabular}{|c|l|c|c|l|c|} 
\hline 
No. &  $~~~~~~~z_{\rm A}$ & ${\rm \log\,N_{\rm A}(H_2)}$ & ${\rm \log\,N_{\rm A}(HD)}$ & ${\rm T_{01},\,K}$ & HD/2H$_2$ \\
\hline
\multirow{2}{*}{\cite{Malec2010}} & $2.0593276(5)$ & $15.76\pm0.03$  & $12.95\pm0.03$  & $116\pm17$  &  $(7.7\pm0.8)\times10^{-4}$\\
                   & $2.0593290(4)$ & $17.55\pm0.04$  & $13.69\pm0.05$  & $-97\pm51$  &  $(6.9\pm1.0)\times10^{-5}$\\
\hline                   
\cite{Tumlinson2010}& $2.0594(1)$ & $17.64\pm0.15$ & $13.84\pm0.20$  & $281^{+2700}_{-134}$  &  $(7.9\pm4.6)\times10^{-5}$\\
\hline
This work & $2.0593245(5)$ & $17.94\pm0.01$ & $13.87\pm0.06$  & $138.6\pm5.8$  &  $(4.3\pm0.6)\times10^{-5}$\\
\hline
\end{tabular}
\end{center}
\end{table*}

\subsection{Analysis of the H$_2$ System in the VLT Spectrum}
\label{H2inVLT}
\begin{figure*}
\begin{center}
        \includegraphics[width=1.0\textwidth]{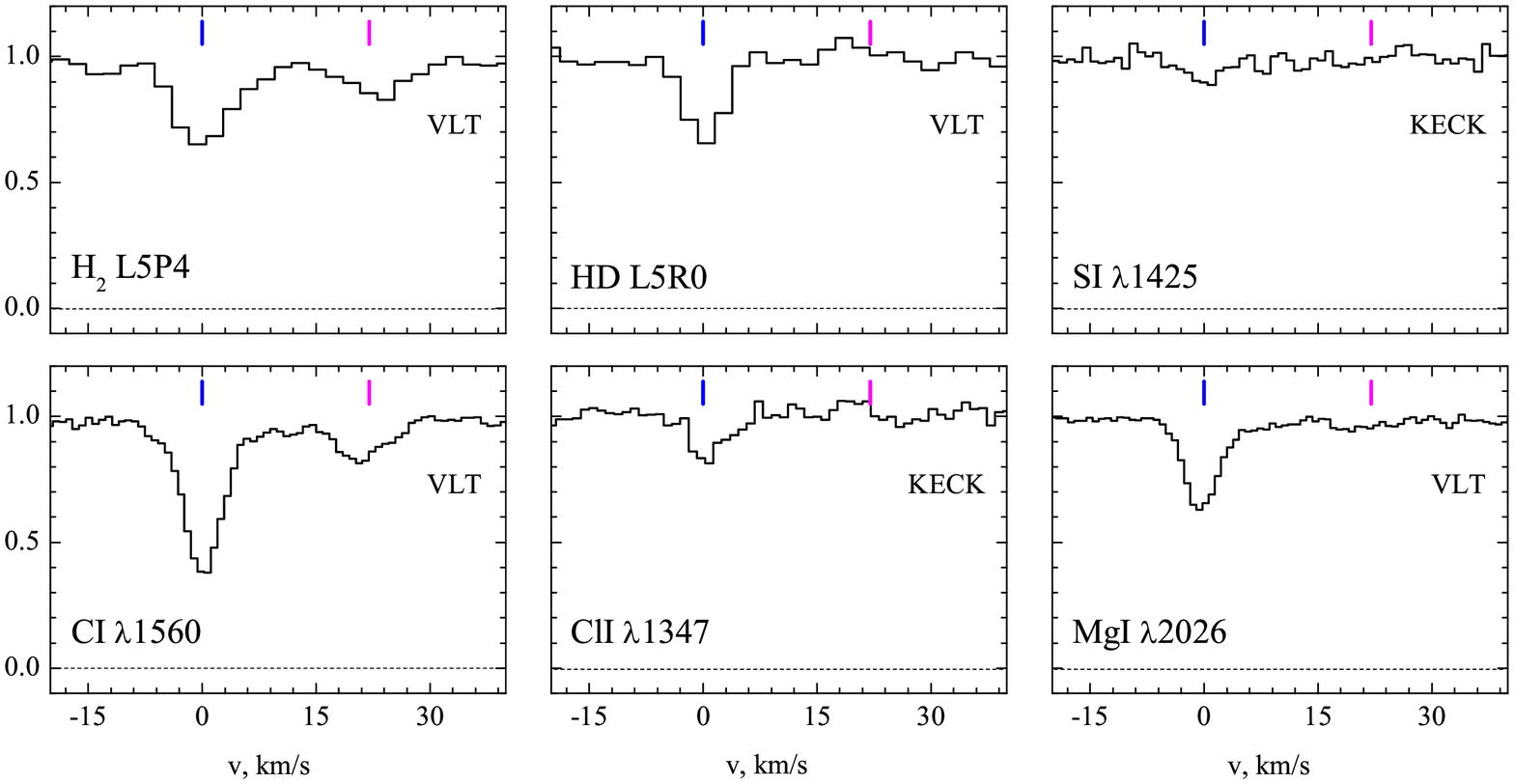}
        \caption{Absorption lines of neutral species associated with theH2 system at $z_{\rm abs} = 2.059$ in the VLT/UVES and Keck/HIRES spectra of the quasar J\,2123$-$0050. The vertical dashes indicate the positions of the two components of the H$_2$ absorption system (A and B) at redshifts $z_{\rm abs} = 2.05933$ and $2.05955$. Only the C\,{\sc i} lines are detected at the redshift of component B. The
velocity shift shown along the horizontal axis is calculated relative to component A.}
        \label{Neutral_el} 
\end{center}
\end{figure*}

\begin{figure*}
\begin{center}
        \includegraphics[width=1.0\textwidth]{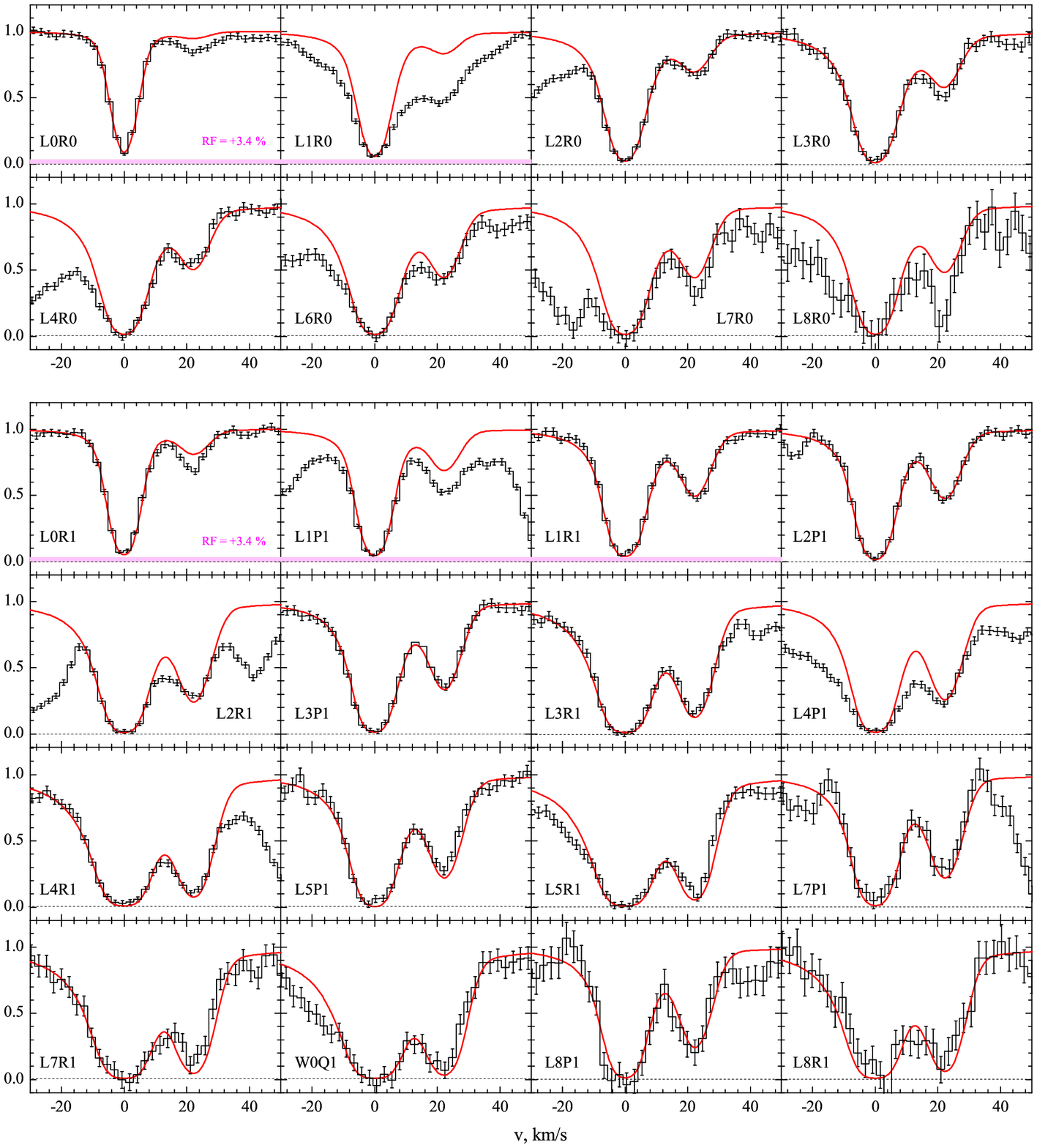}
        \caption{Synthetic spectrum of the H$_2$ system at $z_{\rm abs} = 2.059$ in the spectrum of the quasar J\,2123$-$0050 (VLT/UVES). The lines of the transitions in the H$_2$ absorption system from the rotational ${\rm J = 0, 1}$ levels are shown. The synthetic spectrum was corrected for the coverage factor (see Section\,\ref{part_cov}). The velocity shift shown along the horizontal axis is calculated
relative to component A.}
        \label{H2fit} 
\end{center}
\end{figure*}

The H$_2$ lines of nine Lyman bands and one Werner band fall within the wavelength range of the UVES spectrograph (3000\,\AA\,$<\lambda<11\,000$\,\AA)\footnote{http://www.eso.org/sci/facilities/paranal/instruments/uves/inst.html}. At least two components (hereafter A and B) at redshifts $z_{\rm A} = 2.05932(5)$ and $z_{\rm B} = 2.05955(3)$ are detected in the H$_2$ lines of the transitions from the rotational levels J = 0 to J = 5. The relative velocity shift between components A and B is fairly large
compared to the width of the UVES point spread function $\sim6$\,km\,s$^{-1}$ and the Doppler line broadening $\sim1-5$\,km\,s$^{-1}$ and is $\sim22$\,km\,s$^{-1}$. Therefore, the individual components in the transition lines profiles are resolvable even for the saturated J = 0 and 1 levels. Since no evidence for the presence of additional components was found (see Fig.\,\ref{Neutral_el}), we used a two-component model. The best-fit parameters and their errors are given in Table\,\ref{H2_results}. The synthetic spectrum of the H$_2$ absorption system is shown in Fig.\,\ref{H2fit}. The total column densities are $\log N^{\rm A}_{\rm H_2} = 17.94\pm0.01$ and $\log N^{\rm B}_{\rm H_2}= 15.16\pm0.02$. Note that the Doppler parameter $b_{\rm J}$ increases with increasing level J. This effect is well known for H$_2$ systems \cite{Noterdaeme2007a, Ivanchik2010, Albornoz2014} and can be caused by a nonuniform distribution of H$_2$ molecules at different rotational levels J over the molecular cloud volume (for more details, see \cite{Balashev2009}).

\begin{figure}
\begin{center}
        \includegraphics[width=0.5\textwidth]{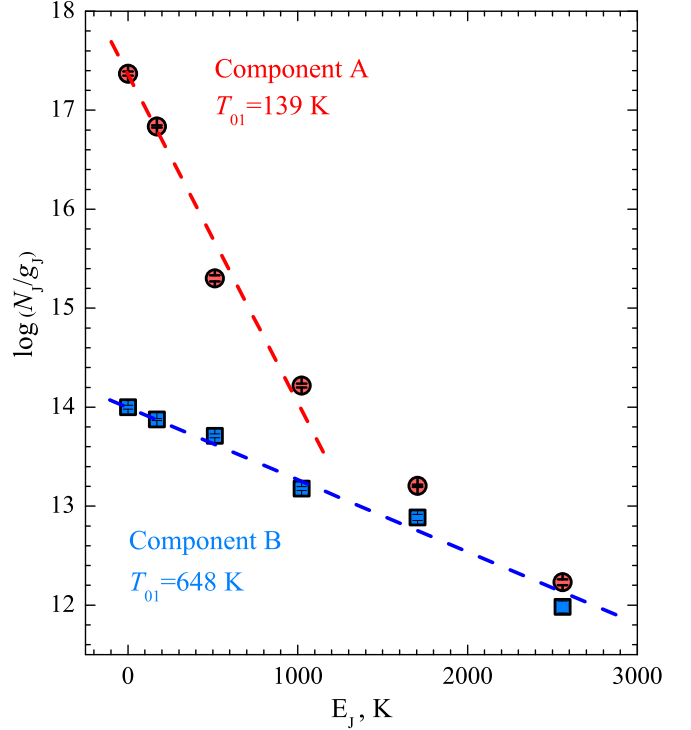}
        \caption{Populations $N_{\rm J}$ of H$_2$ rotational levels normalized to the statistical weight $g_{\rm J}$ versus level excitation energy $E_{\rm J}$. The circles and squares correspond to components A and B of the H$_2$ absorption system. The straight lines correspond to the excitation temperatures $T_{\rm 01}$ determined from the relative populations of the J = 0 and 1 levels.}
        \label{H2exc} 
\end{center}
\end{figure}

In Fig.\,\ref{H2exc} the relative rotational level populations $N_{\rm J}/g_{\rm J}$ are plotted against the level excitation energy $E_{\rm J}$ ($g_{\rm J}$ is the statistical weight of level J) for the two components of the H$_2$ system. For the component A, the J = 0 and 1 levels are populated through collisions.
The kinetic temperature of the gas determined from the ratio of the ortho- and para-hydrogen column
densities is $T_{\rm A} = 139\pm6$\,K. The J = 3, 4, and 5 levels are populated mainly through radiative pumping. In J\,2123$-$0050\,B, the total H$_2$ column density is lower than that in component A by almost three orders of magnitude. In an optically thin medium, the radiative pumping process is important for all H$_2$ levels; therefore, the temperature determined from the ratio of the ortho- and para-hydrogen column
densities, $T_{\rm B}=648\pm126$\,K, can differ significantly from the kinetic temperature of the gas.

\begin{table*}
\begin{center}
\caption{Results of our analysis of the H$_2$ absorption system in the spectrum of J\,2123$-$0050 in the sub-DLA system at $z_{\rm abs}=2.059$.} 
\label{H2_results}
\begin{tabular}{|c|c|c|c|c|c|c|c|} 
\hline \hline
\multicolumn{4}{|c|}{Component A}& \multicolumn{4}{|c|}{ Component B}\\
\multicolumn{4}{|c|}{${\rm z_A = 2.0593245}$}& \multicolumn{4}{|c|}{${\rm z_{B} = 2.0595532}$}\\ 
\hline
~~ &  J & $\log\,N_{\rm A}$ & b$_{\rm A}$, км/с & ~~ & J & $\log\,N_{\rm B}$ & b$_{\rm B}$, км/с \\
\hline
H$_2$ & 0  & $17.37\pm0.02$  & $1.71\pm0.04$  & H$_2$ & 0  & $14.00\pm0.02$ & $5.18\pm0.51$\\
~~    & 1  & $17.79\pm0.01$  & $2.07\pm0.05$  &  ~~   & 1  & $14.84\pm0.01$ & $5.24\pm0.11$\\ 
~~    & 2  & $16.00\pm0.03$  & $2.97\pm0.05$  &  ~~   & 2  & $14.41\pm0.01$ & $5.02\pm0.18$\\ 
~~    & 3  & $15.54\pm0.02$  & $3.56\pm0.06$  &  ~~   & 3  & $14.50\pm0.01$ & $4.71\pm0.30$\\ 
~~    & 4  & $14.16\pm0.01$  & $4.64\pm0.23$  &  ~~   & 4  & $13.84\pm0.03$ & $7.86\pm0.57$\\
~~    & 5  & $13.75\pm0.03$  & $5.10\pm0.51$  &  ~~   & 5  & $13.50\pm0.06$ & $7.36\pm0.80$\\
~~    & $\Sigma_{\rm J}$  & $17.94\pm0.01$  & ~~    & ~~    & $\Sigma_{\rm J}$  & $15.16\pm0.02$& \\ 
\hline
HD    & 0  & $13.87\pm0.06$  & $2.0\pm0.6$    & HD    & 0  & \multicolumn{2}{|c|}{not detected}\\
\hline
HD/2H$_2$    &  \multicolumn{3}{|c|}{$(4.26\pm0.60)\times10^{-5}$}    & HD/2H$_2$    & \multicolumn{3}{|c|}{ ~~}\\
\hline
\end{tabular}
\end{center}
\end{table*}

\subsection{The Residual Flux in Molecular Hydrogen Lines}
\label{part_cov}

For H$_2$ clouds at high redshifts, the partial coverage effect, when the H$_2$ cloud partially covers the
quasar’s emission regions, is possible (see, e.g., \cite{Balashev2011, Albornoz2014, Klimenko2015}). Part of the quasar’s emission passes by the absorption cloud and produces a residual flux (RF) in the spectrum at the bottom of the H$_2$ absorption lines. Disregarding this effect in the analysis of H$_2$ lines can lead to an underestimation of the H$_2$ column density (by up to two orders of magnitude). As it was shown by \cite{Ofengeim2015} the effect of partial coverage of quasars by H$_2$ absorption systems at high redshifts should be observed at least in 10\,\% of the cases.

\begin{figure*}
\begin{center}
        \includegraphics[width=1.0\textwidth]{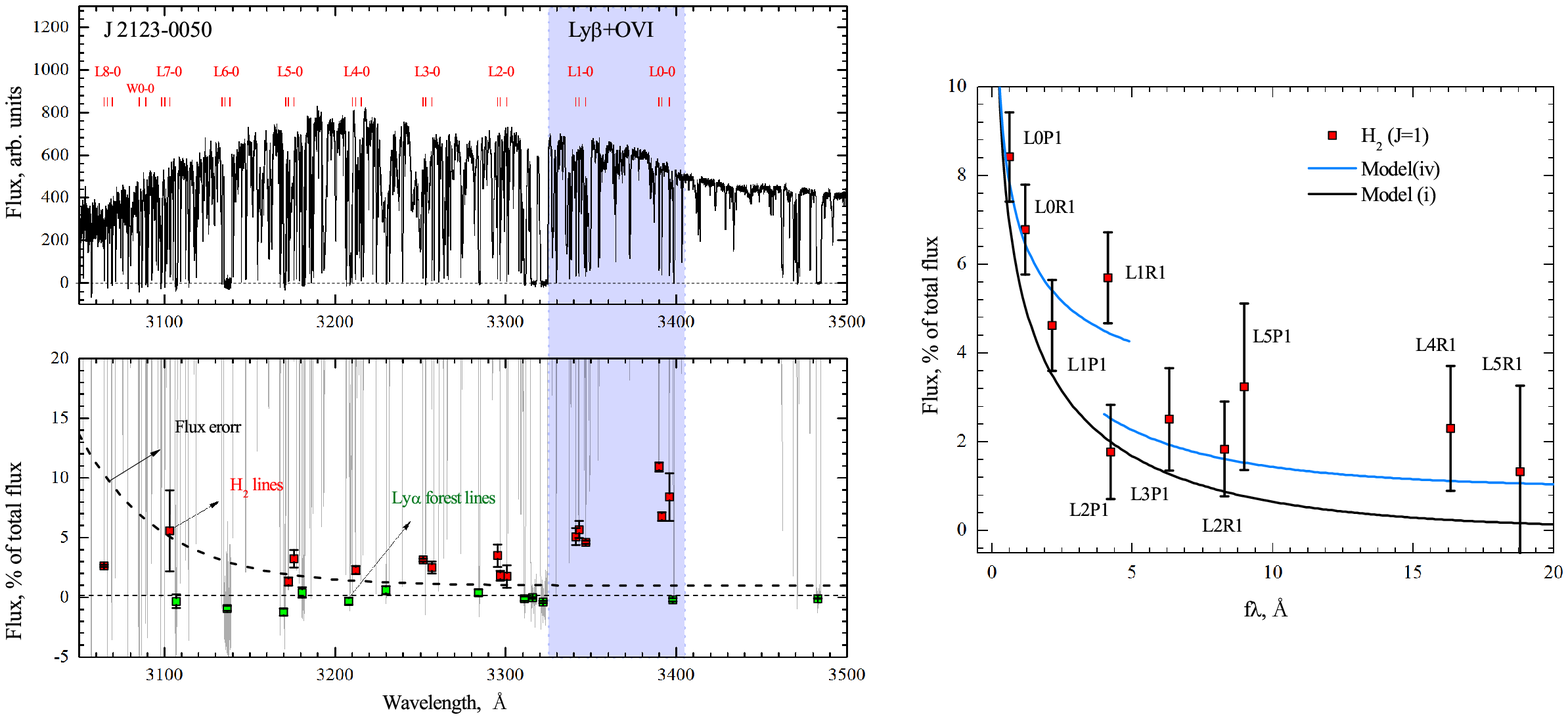}
        \caption{The left panels show the region of the VLT/EVES spectrum for the quasar J 2123.0050 containing the H2 absorption lines. On the upper left panel, the observed flux from the quasar in arbitrary units is shown along the vertical axis; the normalized flux in \% of the total flux is shown on the lower left panel. The gray color highlights the region of the spectrum containing the quasar’s Lyв and OVI emission lines. The open and black filled squares indicate the residual fluxes at the bottom of the H2
and Lyб-forest lines, respectively. The dashed line indicates the statistical error of the flux in the spectrum. The right panel shows the residual fluxes in the H2 lines of the transitions from the J = 1 level as a function of fл, the product of the oscillator strength by the transition wavelength. The solid and dashed lines indicate the calculations performed for the models with and without allowance for the partial coverage effect, models (iv) and (i), respectively ({\bf see Table 3}).}
        \label{H2_res_flux} 
\end{center}
\end{figure*}

In the spectrum of J\,2123$-$0050, we detect a nonzero RF for component A (in the most saturated H$_2$ lines of the transitions from J = 0, 1) in the region of the Ly$\beta$ and O\,{\sc vi} emission lines, $3.4\pm0.5$\,\% of the continuum flux.  Figure\,\ref{H2_res_flux} shows the line positions in the spectrum (upper left panel) and the RF at the line minimum (lower left panel). Analysis of the dependence of the RF on the product of the wavelength and oscillator strength for the H$_2$ transitions (for more details, see \cite{Klimenko2015}) suggests that a nonzero RF in the H$_2$ line is most likely not caused by the convolution of the saturated lines with the instrumental function of the spectrograph but is probably
a consequence of the partial coverage effect (see the right panel in Fig.\,\ref{H2_res_flux}).
To analyze the H$_2$ absorption system, we used four models: (i) without any sources of additional
radiation, (ii) with a source of additional radiation in continuum, (iii) with a source of additional radiation in emission lines, and (iv) with two sources of additional radiation in continuum and emission lines. For a quantitative comparison, we used the statistical AICC (corrected Akaike information criterion; see \cite{Sugiura1978}) criterion\footnote{The most preferred model is determined using the difference
of the AICC values. $\Delta{\rm AICC} = 10$ is deemed to mean strong evidence for the model with the smaller AICC value \cite{Liddle2007}.}. The results of our analysis are presented in Table\,\ref{H2_compare}. For models (iii) and (iv), the AICC values are found to be lower ($\Delta{\rm AICC} \sim -40$) than those for models (i) and (ii). Taking into account the RF in continuum allows the AICC values to be reduced by 10 more units. The H2 lines are best described by model (iv). It should be noted that in view of the low H$_2$ column density and the RF smallness (about $1-4$\,\%), allowance for the partial coverage effect affects weakly
the H$_2$ column densities being determined (see columns 7 and 8 in Table\,\ref{H2_compare}). However, this case increases the statistics of observations of the partial coverage effect for high-redshift H$_2$ absorption systems
to four cases.

\section{ANALYSIS OF HD LINES}

\begin{figure*}
\begin{center}
        \includegraphics[width=1.0\textwidth]{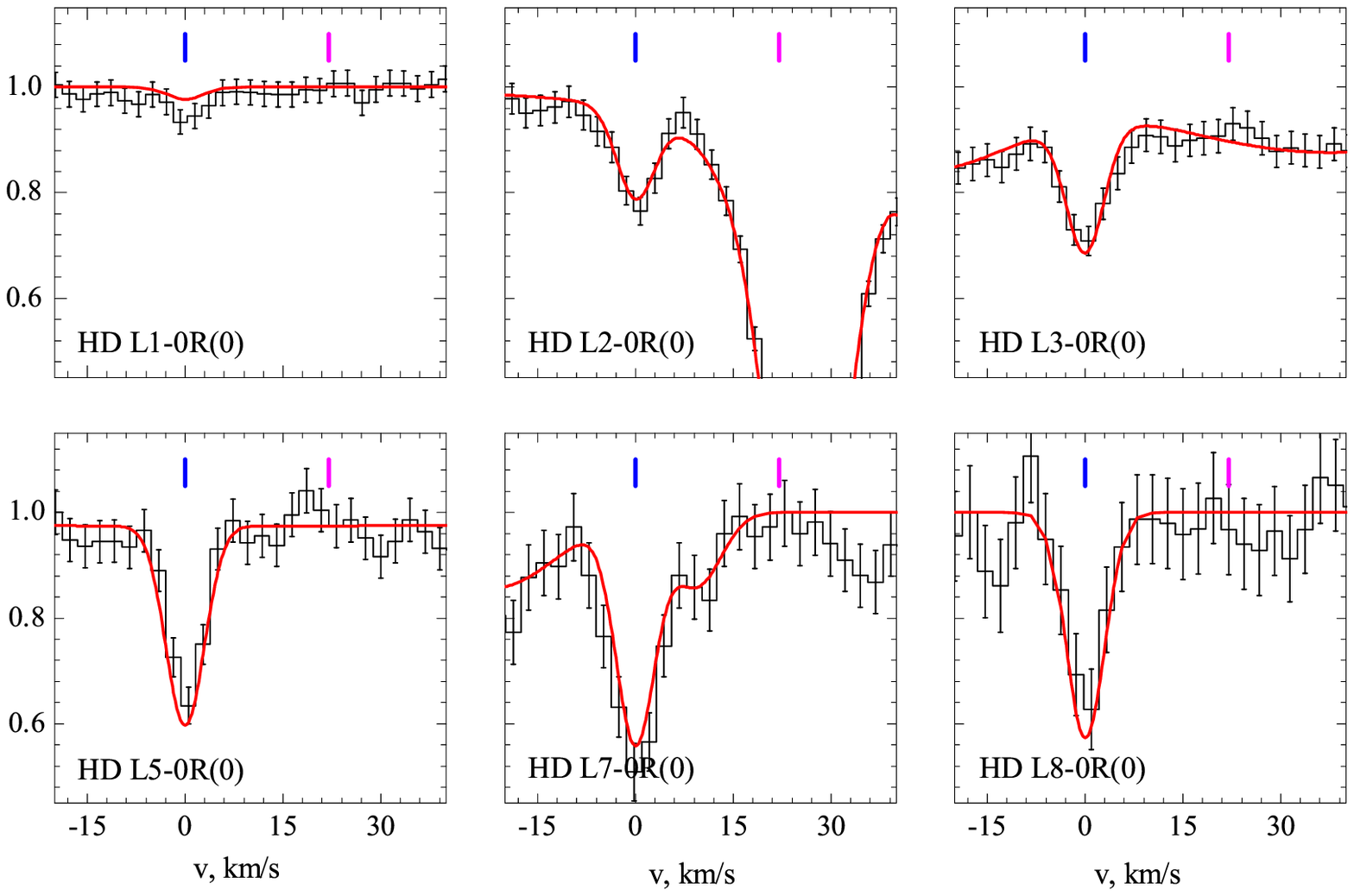}
        \caption{Synthetic spectrum of HD molecules at $z_{\rm abs} = 2.05933$ fitted into the observed spectrum of the quasar J\,2123$-$0050 (VLT/UVES). The vertical dashes indicate the positions of the two components of the H$_2$ absorption system (A and B). The velocity shift shown along the horizontal axis is calculated relative to component A, in which the HD lines are detected.}
        \label{HDlines} 
\end{center}
\end{figure*}

To estimate the column density of HD molecules, we used the VLT spectrum. Because of the smallness
of the H2 column density in component B ($\log N^{\rm B}_{\rm H_2}=15.16\pm0.02$), the expected column density of HD molecules does not exceed $\log N^{\rm B}_{\rm HD}\le11$; therefore, it is impossible to detect the HD lines in this component at the existing sensitivity level of modern telescopes.
The lines of HD molecules are detected only in component A for the transitions of the Lyman and
Werner bands up to L8-0 and W0-0. To determine the HD column density (${\rm J = 0}$), we used the L0-0,
L2-0, L3-0, L5-0, L7-0, and L8-0 transition lines of HD molecules, which do not overlap with the Ly$\alpha$-forest lines. The laboratory wavelengths of the HD transitions were taken from \cite{Ivanov2010}. The oscillator strengths were taken from the calculations by \cite{Abgrall2006}. The estimated column density of HD molecules is $\log N^{\rm A}_{\rm HD} = 13.87\pm0.06$. The HD line profiles and the fitted synthetic spectrum are shown in Fig.\,\ref{HDlines}. The derived value is consistent with the estimate $\log N^{\rm A}_{\rm HD} = 13.84\pm0.20$ \cite{Tumlinson2010}. However, owing to the higher signal-to-noise ratio in the VLT spectrum, our estimate of $N_{\rm HD}$ is considerably more accurate.

\begin{table*}
\begin{center}
\caption{Comparison of the results of our analysis of the H$_2$ absorption system for various residual flux (RF) models. The columns present the following: 2 -- the reduced $\chi^2$ value; 3 -- the absolute AICC value \cite{Sugiura1978}; 4 -- the difference of the AICC values; 5 and 6 -- RF$_{\rm cont}$ and RF$_{\rm emis}$ -- the RFs due to the effect of partial coverage of the quasar’s emission region in continuum
and emission lines, respectively; 7 and 8 -- the measured H$_2$ column densities in component A for the transitions from the J = 0and 1 levels.} 
\label{H2_compare}
\begin{tabular}{|c|c|c|c|c|c|c|c|} 
\hline \hline
Model &   $\chi^2_{red}$ & AICC & ${\rm{\Delta AICC}}$ & $\mbox{RF}_{\mbox{cont}}$ & $\mbox{RF}_{{\mbox{emis}}}$ & $\log\,N^{\rm A}_{\rm H_2}({\rm J=0})$ & $\log\,N^{\rm A}_{\rm H_2}({\rm J=1})$  \\
\hline
i        &  1.11   & 374.7 &  0      &   no                &   no        & $17.36\pm0.02$ & $17.79\pm0.01$ \\
ii       &  1.05   & 363.9 &  -10.8  &   $1.1\pm0.4$       &   no        & $17.38\pm0.02$ & $17.80\pm0.01$ \\
iii      &  0.94   & 330.0 &  -44.7  &   no                & $2.7\pm0.5$ & $17.35\pm0.02$ & $17.78\pm0.01$ \\
iv       &  0.91   & 321.3 &  -53.4  &   $1.1\pm0.3$       & $3.4\pm0.5$ & $17.37\pm0.02$ & $17.79\pm0.01$ \\
\hline
\end{tabular}
\end{center}
\end{table*}   

\begin{table*}
\begin{center}
\caption{Comparison of the results of analysis of the C\,{\sc i} lines in the spectrum of J\,2123$-$0050 for four models. (i) and (ii) -- without allowance for the RF in the C\,{\sc iv} emission line, (iii) and (iv) -- with allowance for the RF. The RF is measured in\,\% of the total flux. The redshifts of the components are: $z_{\rm A}=2.0593245(4)$, $z_{\rm B}=2.059546(3)$, $z_{\rm C}=2.059330(3)$. 
} \label{table_carbon}
\begin{tabular}{|c|c|c|c|c|c|c||c|c|} 
\hline
\multicolumn{2}{|c|}{Model} & $\log N_{\rm \,C\,I}$ & $\log N_{\rm \,C\,I^*}$ & $\log N_{\rm \,C\,I^{**}}$ & $b$, km\,s$^{-1}$ & RF & $\chi^2_{\rm red}$ & AICC\\ 
\hline
\multirow{2}{*}{(i)} & A & $13.82\pm0.02$ & $13.26\pm0.02$ & $12.65\pm0.02$ & $1.31\pm0.03$ & \multirow{2}{*}{no} & \multirow{2}{*}{1.53} & \multirow{2}{*}{638}\\
 & B & $12.71\pm0.02$ & $12.43\pm0.04$ & $11.94\pm0.18$ & $4.37\pm0.44$ & & & \\
\hline 
\multirow{3}{*}{(ii)}  & A & $13.91\pm0.03$ & $13.46\pm0.03$ & $12.65\pm0.03$ & $0.88\pm0.04$ & \multirow{3}{*}{no} & \multirow{3}{*}{0.91} & \multirow{3}{*}{386.4}\\
 & B & $12.71\pm0.02$ & $12.43\pm0.03$ & $11.90\pm0.09$ & $4.18\pm0.27$ &  & & \\ 
 & C & $12.78\pm0.03$ &  no  &  no  & $6.89\pm0.60$ & & & \\ 
\hline
\multirow{2}{*}{(iii)} & A & $13.84\pm0.03$ & $13.27\pm0.02$ & $12.63\pm0.04$ & $1.32\pm0.04$ & \multirow{2}{*}{$1\pm1$} & \multirow{2}{*}{1.52} & \multirow{2}{*}{635}\\
 & B & $12.71\pm0.03$ & $12.43\pm0.04$ & $11.90\pm0.18$ & $4.40\pm0.50$ & & & \\
\hline 
\multirow{3}{*}{(iv)} & A & $13.91\pm0.03$ & $13.47\pm0.04$ & $12.64\pm0.03$ & $0.87\pm0.04$ & \multirow{3}{*}{$1\pm1$} & \multirow{3}{*}{0.91} & \multirow{3}{*}{388.4}\\
 & B & $12.71\pm0.02$ & $12.43\pm0.03$ & $11.85\pm0.13$ & $4.25\pm0.27$ &  & & \\
 & C & $12.78\pm0.03$ &  no  &  no  & $7.26\pm0.70$  &  & & \\ 
\hline 
\end{tabular}
\end{center}
\end{table*}

\section{NEUTRAL CARBON}

The structure of the neutral carbon line for the  transitions from the ground, ${\rm 2s^22p^{23}P_{0}}$ (C\,{\sc i}), and two excited, ${\rm 2s^22p^{23}P_{1}}$ (C\,{\sc i}$^*$) and ${\rm 2s^22p^{23}P_{2}}$ (C\,{\sc i}$^{**}$), states consists of at least two components at redshifts coincident with the positions of components A and B of the H$_2$ system. The neutral carbon lines were analyzed for four models. Model (i) consists of two C\,{\sc i} components associated with components A and B of the H$_2$ absorption system. In comparison with model (i), an additional component with a considerably larger Doppler parameter is used in model (ii). Models (iii) and (iv) differ from models (i) and (ii) by allowance for the RF for the C\,{\sc i} lines falling into the C\,{\sc iv} emission line wing (for more details, see Section\,\ref{part_cov_ci}). The results of our analysis are presented  in Table\,\ref{table_carbon}.

\begin{figure*}
\begin{center}
        \includegraphics[width=0.9\textwidth]{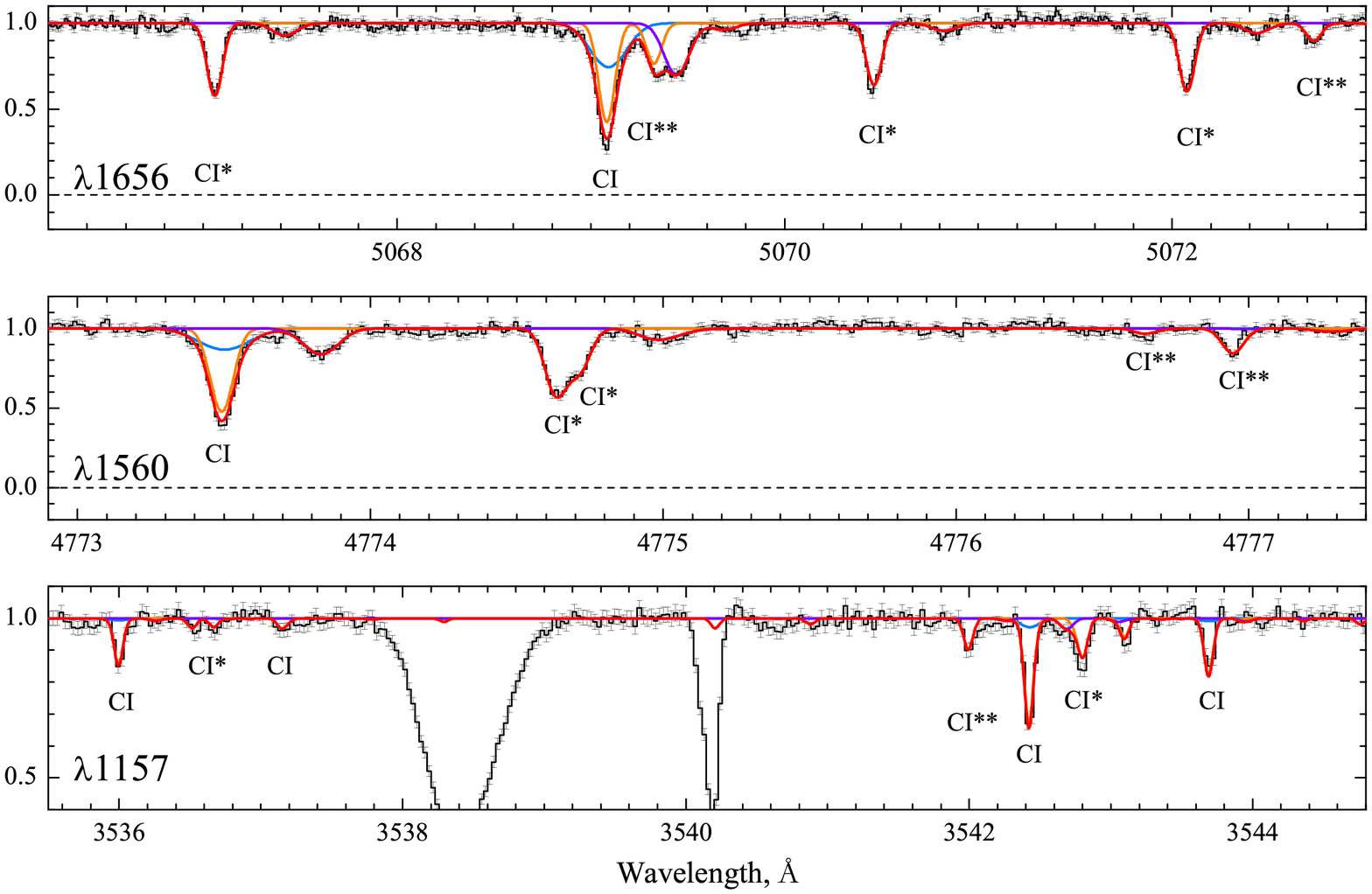}
        \caption{Synthetic spectrum of the C\,{\sc i} lines associated with the H$_2$ absorption system fitted into the observed spectrum of the quasar J\,2123$-$0050 (UVES/VLT). The profiles of the individual C\,{\sc i} line components are indicated by the thin solid lines.}
        \label{Cilines} 
\end{center}
\end{figure*}

Model (i) describes the C\,{\sc i} line profiles for the transitions from the ground state for component A
not well enough ($\chi^2_{\rm red}\approx1.5$). Model (ii) gives a considerably better description ($\chi^2_{\rm red}\approx0.9$) the change in the AICC value compared to model (i) is $\Delta{\rm AICC}=-264$, which, according to the estimate by \cite{Liddle2007}, is strong evidence for model (ii).
Figure\,{\ref{Cilines}}  shows the synthetic spectrum for model (ii) fitted into the quasar’s observed spectrum. The appearance of an additional C\,{\sc i} subcomponent that is not detected in the H$_2$ lines can be explained by the presence of neutral carbon associated with the ionized part of the sub-DLA system.

According to the estimate by \cite{Milutinovic2010}, up to 90\,\% of the matter in the sub-DLA
system being investigated is in an ionized state. Assuming that the C$^+$ ions are distributed uniformly
over the volume, the bulk of the C$^+$ column density belongs to the ionized gas, where the electron number
density ($n_{\rm e}\sim n_{\rm H^+}$) turns out to be three orders of magnitude higher than that in the neutral medium ($n_{\rm e} < 10^{-3}\times n_{\rm H}$; see e.g., \cite{Dalgarno1984, Welty2003}). As a
result of these two factors, the amount of C\,{\sc i} formed in the recombination reaction of C$^+$ ions with electrons can reach the observed value in component C: $\log N_{\rm C\,I} \sim 13$. From the chemical equilibrium condition,
\begin{equation}
\label{eqCI}
N_{\rm C\,I}=N_{\rm C^+}\frac{n_{\rm e} \alpha({\rm C^+})}{\rm \beta{\rm(C)}}, 
\end{equation} 
where $\alpha(\rm C^+)$ is the recombination coefficient ({$\alpha(\rm C^+)\sim10^{-12}$}\,cm$^3$s$^{-1}$ for $T\simeq3\times10^3$\,K, see \cite{Nahar1997}), $\beta{\rm(C)}=\chi_{\rm UV}\times2.1\times10^{-10}\,\mbox{s}^{-1}$ -- is the CI photoionization rate  (see., e.g., \cite{LePetit2002}), $\chi_{\rm UV}$ is the UV background intensity with respect to the mean Galactic value \cite{Habing1968}. However, since the C\,{\sc ii} 1334.5\,\AA\, absorption line in the spectrum consists of several overlapping saturated components, the total C$^+$ column density cannot be determined. On the other hand, the amount of C$^+$ can be estimated by assuming that the carbon abundance with respect to its solar abundance is the same as that for sulfur (S). For metallicity ${\rm [S/H]=-0.2}$  and $\log N_{\rm H}^{\rm tot}=20.06$ \cite{Milutinovic2010}, we obtain $\log N_{\rm C^+}^{\rm tot}=16.25$. The fraction of C$^+$ ions associated with the ionized part of the sub-DLA system can be estimated as  $ N_{\rm C^+}^{\rm tot}\times N_{\rm H^+}/ N_{\rm H}^{\rm tot}=10^{16.2}$. Then, according to Eq.\,\eqref{eqCI}, we obtain

\begin{equation}
N_{\rm C\,I}=0.8\times10^{13}\times\frac{n_{\rm e}}{\mbox{1\,см$^{-3}$}}\times\frac{10}{\chi_{\rm UV}}\,\mbox{cm}^{-2}, 
\end{equation}  

which is consistent with $\log N^{\rm C}_{\rm C\,I}=12.78$ for model (ii) from Table\,\ref{table_carbon}.

\subsection{The Partial Coverage Effect for the Neutral Carbon System}
\label{part_cov_ci}

The absorption lines of two transitions, C\,{\sc i} 1328.8\,\AA\, and 1656.9\,\AA,\,, shifted by the factor $(1 + z_{\rm abs})$ are in the region of the quasar  N\,{\sc v} и C\,{\sc iv} emission lines, respectively. Out of these C\,{\sc i}  lines, only the C\,{\sc i} 1656.9\,\AA\, line falls within the wavelength range of the quasar's VLT spectrum accessible to analysis. As was shown by \cite{Balashev2011}, when analyzing
the quasar Q\,1232$+$082, the residual flux in the C\,{\sc i} lines due to the partial coverage effect can reach $20-30$\,\% of the total flux, which can change significantly the C\,{\sc i} column densities being determined. To check this possibility, we varied the residual flux in the C\,{\sc i} lines near the C\,{\sc iv} emission line, along with other parameters. The results of our analysis with (models (iii) and (iv)) and without (models (i) and (ii)) taking into account the RF are compared in Table\,\ref{table_carbon}. In models (ii) and (iv), the RF is detected at $1\pm1$\,\% of the total flux. However, since the C\,{\sc i} lines are unsaturated, allowance for the RF in our analysis barely changes the $\chi^2$ and AICC values (compared to those for models (i) and (iii), respectively). Thus, we cannot unambiguously determine whether the partial coverage effect is present or absent in the C\,{\sc i} lines (at a $1\pm1$\,\% level) for this system.

\section{COMPARISON OF OUR RESULTS WITH THE RESULTS OF PREVIOUS ANALYSES OF THIS SYSTEM}

\subsection{The H$_2$ Column Density in the Component A}

\begin{figure*}
\begin{center}
        \includegraphics[width=1.0\textwidth]{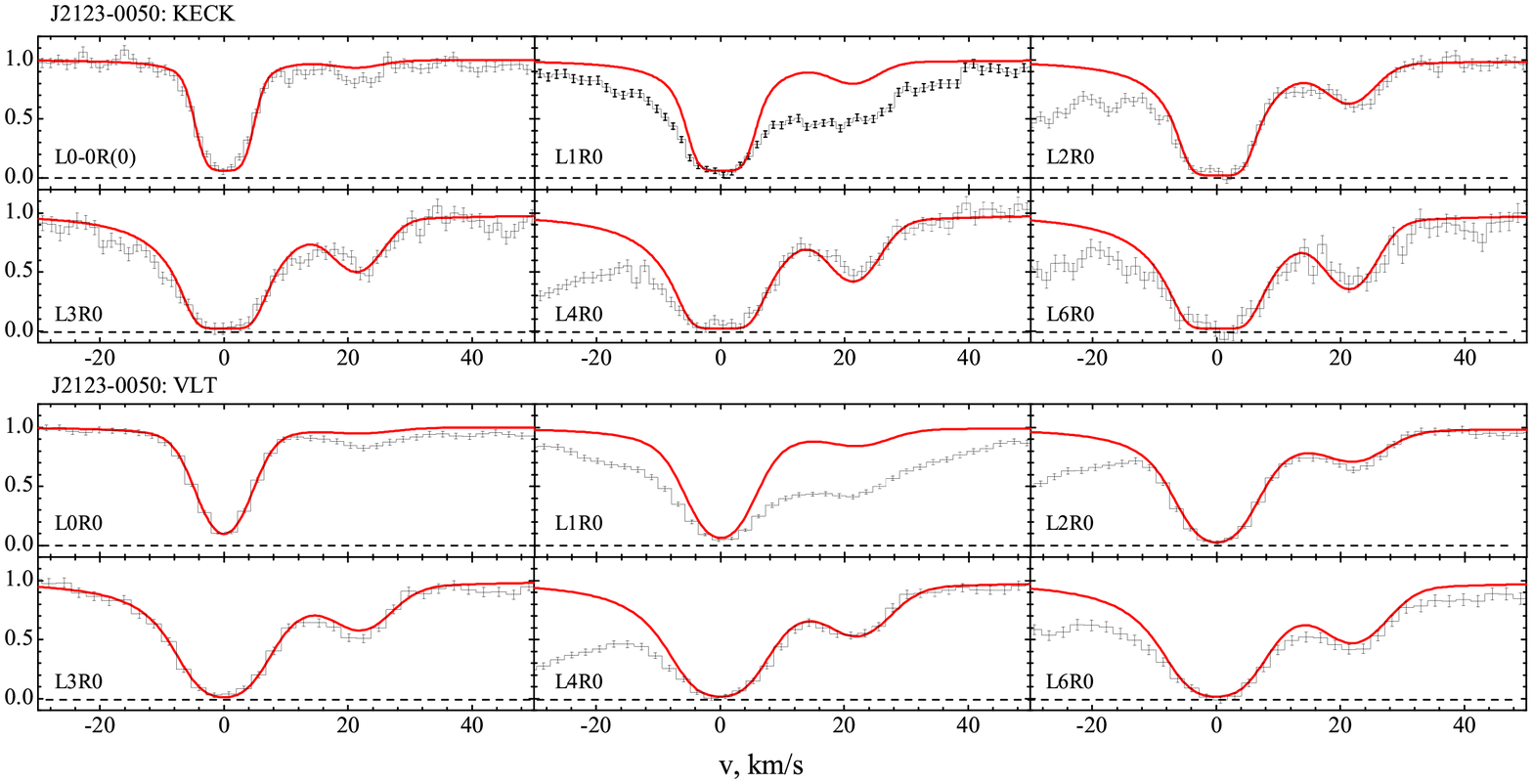}
        \caption{Synthetic spectrum of the H$_2$ lines of the transitions from the J = 0 level fitted into the Keck/HIRES (upper panel) and VLT/UVES (lower panel) spectra of the quasar J\,2123$-$0050. The H$_2$ column density in component A was assumed to be $\log N^{\rm A}_{\rm H_2}{\rm (J = 0) = 17.37}$. Since the widths of the HIRES/Keck ($\sim3$\,km/s) and UVES/VLT ($\sim6$\,km/s) point spread functions differ, the profiles of the corresponding H$_2$ lines on the upper and lower panels differ.}
        \label{H2j01vltkeck} 
\end{center}
\end{figure*}

The results of our analysis of the H$_2$ system in the VLT spectrum of J\,2123$-$0050 are compared with the
results of the analyses of the H$_2$ system in the Keck spectrum of J\,2123$-$0050 \cite{Malec2010, Tumlinson2010} in Table\,\ref{table_1}. The results differ mainly in H$_2$ column density estimate for component A at the J = 0 level. $N^{\rm A}_{\rm H_2}({\rm J = 0})$ obtained in this paper is $\log N^{\rm A}_{\rm H_2}({\rm J = 0})=17.37\pm0.02$, which exceeds the estimate from \cite{Tumlinson2010} by a factor of 3 $(16.86\pm0.24)$ and the estimate from \cite{Malec2010} by a factor of 30 ($15.80\pm0.40$). Different quality of the spectra and different allowance for the Ly$\alpha$-forest absorption lines overlapping with the H$_2$ lines can be responsible for the difference between the results of our analysis and those of \cite{Malec2010, Tumlinson2010}. Among all the H$_2$ lines of the transitions from the J = 0 level accessible to analysis, only two lines, L0R0 and L3R0, do not overlap with the Lyб-forest lines. Figure\,\ref{H2j01vltkeck} shows the profiles for the H$_2$ absorption lines of the transitions from the J = 0 level for the Keck (upper panel) and VLT (lower panel) spectra. The solid curve indicates the synthetic spectrum of the H$_2$ system constructed using the H$_2$ column densities determined in this paper. The Lyб-forest lines in the profiles for the H$_2$ lines of the L2R0 and L4R0 transitions are seen to fall into the H$_2$ line wing, while the Ly$\alpha$-forest lines for the L1R0 and L6R0 transitions are located near the H$_2$ line center. To determine the H$_2$ column density at the J = 0 level by comparing the observed spectrum
with the synthetic one using all of the accessible H$_2$ lines, it is necessary to artificially add the Ly$\alpha$-forest absorption lines to the synthetic spectrum of the H$_2$ system. The spectral parameters ($N$ and $b$) and the number of components for the unsaturated Ly$\alpha$-forest lines cannot be unambiguously determined in most cases, because such systems belong to the root part of the curve of growth. Therefore, disregarding or, conversely, overusing the Ly$\alpha$-forest lines in the analysis can lead to an overestimation or underestimation of the H$_2$ column density being determined, respectively.

\subsection{The HD/2H$_2$ ratio}

\begin{figure*}
\begin{center}
        \includegraphics[width=0.8\textwidth]{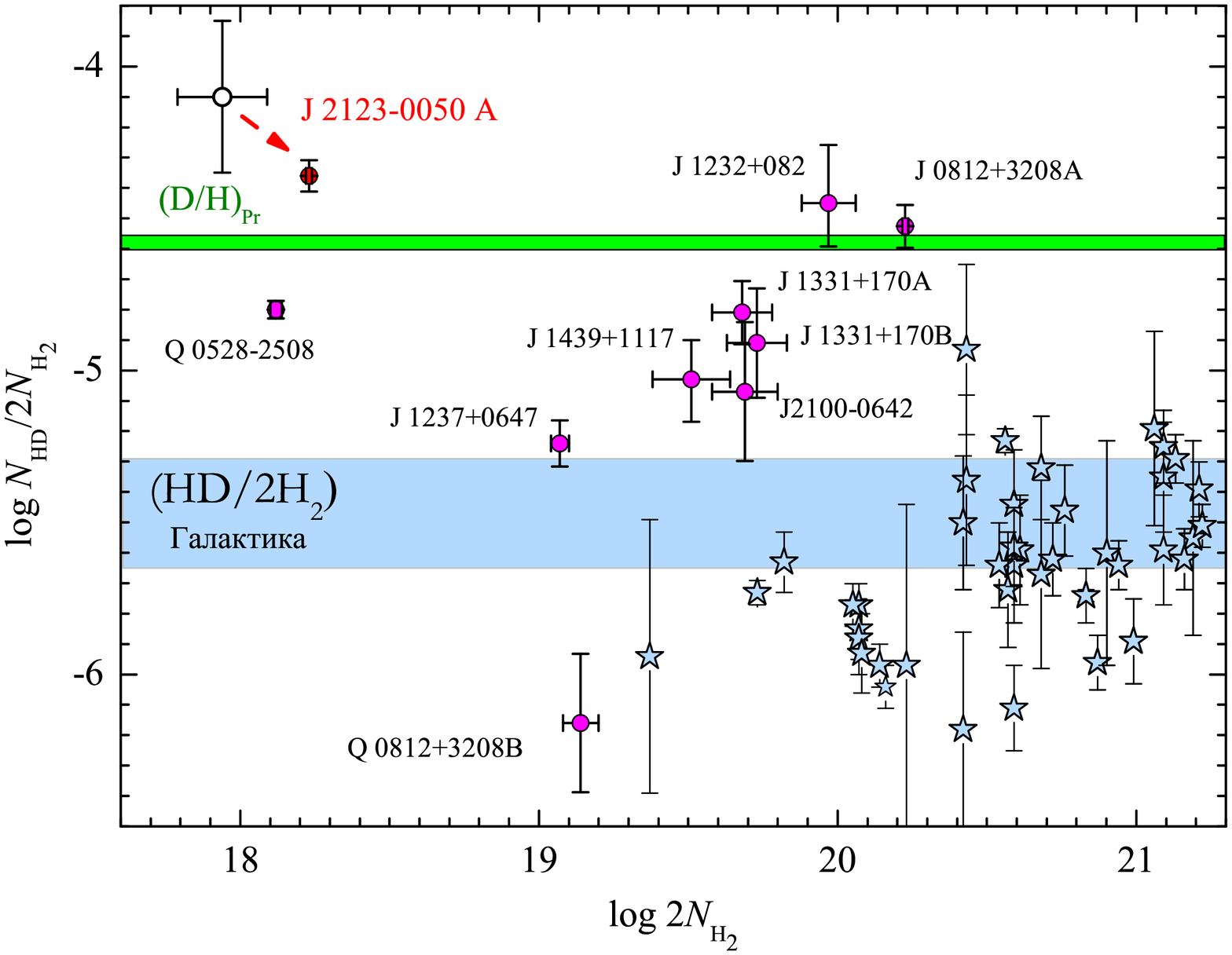}
        \caption{Measured $N_{\rm HD}/2N_{\rm H_2}$ ratios for high-redshift absorption systems (filled circles) and for systems in our Galaxy (open stars). The solid horizontal line indicates the relative primordial deuterium abundance ((D/H)) estimated by analyzing the CMBR anisotropy power spectrum \cite{Planck2015}. The result of the H$_2$/HD system in the spectrum of J\,2123$-$0050 presented in \cite{Tumlinson2010} is indicated by the open circle.}
        \label{hdvsh2} 
\end{center}
\end{figure*}

Using the total H$_2$ and HD column densities in component A, $\log N^{\rm A}_{\rm H_2}= 17.94\pm0.01$ and $\log N^{\rm A}_{\rm HD}= 13.87\pm0.06$, we determined the HD/2H$_2$ ratio for component A, $N^{\rm A}_{\rm HD}/2N^{\rm A}_{\rm H_2}=(4.26\pm0.60)\times10^{-5}$. This value is lower than that measured by \cite{Tumlinson2010}, $(7.9\pm4.6)\times10^{-5}$, but it exceeds the primordial deuterium abundance
estimated by \cite{Planck2015}, $(2.62\pm0.15)\times10^{-5}$, by almost 3 standard deviations. Figure\,\ref{hdvsh2} compares the measured $N_{\rm HD}/2N_{\rm H_2}$ in the absorption systems in the spectra of quasars at $z_{\rm abs} > 2$ and in the systems of our Galaxy. The data were taken from Table\,2 in \cite{Ivanchik2015} (quasars) as well as \cite{Snow2008, Lacour2005} (our Galaxy). At present, the H$_2$/HD system being investigated is the only one in which the $N_{\rm HD}/2N_{\rm H_2}$ ratio exceeds the primordial deuteriumabundance estimate. The possible explanations of such a high value are most likely related to the chemistry of molecular clouds; more specifically, under certain physical  conditions in the cloud, the deuterium molecular fraction can exceeds the hydrogen molecular fraction (\cite{Balashev2010} see also Sect.\ref{PDRsec}).

\section{PHYSICAL CONDITIONS IN THE MOLECULAR CLOUDS J\,2123$-$0050 A AND B}
\label{sec:physcond}
\subsection{The UV Background Intensity}

The upper rotational levels of H$_2$ molecules with J = 3, 4, and 5 are populatedmainly through radiative
pumping (see, e.g., \cite{Black1976}). The H$_2$ rotational level population diagrams for components
A and B are shown in Fig.\,\ref{H2exc}. Component A with a high column density corresponds to the optically
thick case \cite{Jura1975b}, while component B has a low column density and represents the optically
thin case \cite{Jura1975a}. In the optically thin case, the shielding of UV radiation is negligible. This allows a simple estimate of the UV background intensity to be obtained within the framework of a homogeneous
model, where the gas number density and the UV background intensity are assumed to be constant over the cloud volume. In the optically thick case, as the depth of radiation penetration into the cloud increases, the UV background intensity can decrease by several orders of magnitude due to the effect of shielding by H$_2$ molecules (see, e.g., \cite{Abgrall1992, LePetit2006}). At the center of an optically thick cloud, the H$_2$ molecules are predominantly at the lower J = 0 and 1 rotational levels, while the H$_2$ molecules near the cloud boundary can strongly populate the upper J = 3, 4, and 5 rotational levels (see, e.g., \cite{Abgrall1992, Balashev2009}). Therefore, numerical simulations, for example, with the Meudon PDR \cite{LePetit2006} or CLOUDY \cite{Ferland2013} codes, should be used for detailed calculations. Our calculation within the homogeneous model gives only an approximate estimate.

The component B represents the optically thin case that can be described in the approximation of a homogeneous cloud. The ortho- (J = 1, 3, 5) and para- (J = 2, 4, 6) hydrogen level populations are determined by radiative pumping, H$_2$ collisions withHatoms, and spontaneous transitions:
\begin{equation}
\begin{split}
&N_{\rm H_2}^{\rm J=0}\beta p_{\rm 4,0}+N_{\rm H_2}^{\rm J=2}(n_{\rm H} k_{\rm 24}+\beta p_{\rm 4,2}) ={}\\ &=N_{\rm H_2}^{\rm J=4}(A_{\rm 42}+n_{\rm H} k_{\rm42}+ \beta (p_{\rm 2,4}+p_{\rm 0,4}))
\end{split}
\end{equation}
\begin{equation}
\begin{split}
&N_{\rm H_2}^{\rm J=1}\beta p_{\rm 5,0}+N_{\rm H_2}^{\rm J=3}(n_{\rm H} k_{\rm 35}+\beta p_{\rm 5,3}) ={}\\ &=N_{\rm H_2}^{\rm J=5}(A_{\rm 53}+n_{\rm H} k_{\rm 53}+ \beta (p_{\rm 3,5}+p_{\rm 1,5}))
\end{split}
\end{equation}
where $\beta$ is the photoabsorption rate, $k_{\rm 24}$ and $k_{\rm 42}$ are the collisional rate coefficients for the transitions between the H$_2$ levels ($2 \to 4$) and ($4\to2$),
$k_{\rm 35}$ and $k_{\rm 53}$ are those for the transitions between the H$_2$ levels ($3\to5$) and ($5\to3$), $p_{\rm4,0}=0.26$, $p_{\rm2,4}=0.32$, $p_{\rm4,2}=0.32$, $p_{\rm0,4}=0.08$, $p_{\rm5,1}=0.12$ и $p_{\rm5,3}=0.21$ are the radiative pumping rate coefficients \cite{Jura1975a}, and $A_{\rm42}=2.79\times10^{-9}$\,s$^{-1}$ and $A_{\rm53}=9.8\times10^{-9}\,\mbox{s}^{-1}$
 are the spontaneous transition probabilities \cite{Spitzer1978}. For a temperature of $\sim200$\,K,
$k_{\rm 24}=8.8\times10^{-16}$ and $k_{\rm42}=1.8\times10^{-13}$, $k_{\rm35}=6\times10^{-17}$ and $k_{\rm53}=5\times10^{-14}$ in units of cm$^3$\,s$^{-1}$ \cite{Forrey1997}; therefore, for a gas number density $n_{\rm H} < 100$\,cm$^{-3}$, the contribution to the populations of the J = 4 and 5 level through the collisions of H and H$_2$ is negligible. The photoabsorption rate of UV radiation in the cloud can then be estimated as

\begin{equation}
\beta_{\rm J=4}=\frac{N_{\rm H_2}^{\rm J=4} A_{\rm 42}}{N_{\rm H_2}^{\rm J=0} p_{\rm 4,0}+N_{\rm H_2}^{\rm J=2} p_{4,2}},
\end{equation}      

\begin{equation}
\beta_{\rm J=5}=\frac{N_{\rm H_2}^{\rm J=5} A_{\rm 53}}{N_{\rm H_2}^{\rm J=1} p_{\rm 5,1}+N_{\rm H_2}^{\rm J=3} p_{\rm 5,3}}.
\end{equation} 
Using $N_{\rm H_2}^{\rm J}$ from Table.\,\ref{H2_results}, we obtain  $\beta=1.8\times10^{-9}\,\mbox{s}^{-1}$ (for J=4) and $\beta=2.1\times10^{-9}\,\mbox{s}^{-1}$ (for J=5).

The photoabsorption rate $\beta$ is related to the intensity of external UV radiation by the following
relation: $\beta=4\pi\times10^9 J_{\rm UV} S_{\rm shield}(N_{\rm H_2})$, where $J_{\rm UV}\mbox{\,(erg\,s$^{-1}$\,cm$^2$\,Hz$^{-1}$\,rad$^{-1})$}$  is the intensity
of UV radiation with energy $E = 12.87$\,eV averaged over the solid angle \cite{Abel1997}, and
$S_{\rm shield}(N_{\rm H_2})$ is the factor that takes into account the self-shielding of the H$_2$ cloud from the UV background \cite{Draine1996}\footnote{For $\log N_{\rm H_2}>14$: $S_{\rm shield}(N_{\rm H_2})={0.965}/{(1+x/b_5)^2}+{0.035}/{(1+x)^{0.5}}\times\exp(-8.5\times10^{-4}(1+x)^{0.5})$, where $x=N/5\times10^{14}$\,cm$^{-2}$, $b_5=b/10^5$\,cm\,s$^{-1}$}. The mean UV background intensity in our Galaxy is $J^{\rm G}_{\rm UV}\simeq3.2\times10^{-20}\mbox{erg\,s$^{-1}$\,cm$^2$\,Hz$^{-1}$\,rad$^{-1}$}$ (see,
e.g., \cite{Habing1968, Hirashita2005}). Using the H$_2$ column density at the cloud center, $\log N_{\rm H_2}^{\rm B}=14.86$, we find that the intensity of the external UV background in component B is higher
than the mean Galactic value by a factor of $\chi_{\rm UV}=J_{\rm UV}/J^{\rm G}_{\rm UV}=8.3$.

\subsection{The Gas Number Density for Components A and B}

Using the C\,{\sc i} fine-structure level populations, we can estimate the gas number density and temperature
in the cloud and the UV background intensity. In calculating the balance of C\,{\sc i} level populations, we
took into account the interaction of C\,{\sc i} atoms with cosmic microwave background radiation (CMBR)
photons ($T_{\rm CMB} = 2.725\times(1+z) = 8.34$\,K), the radiative pumping by UV radiation, and the collisions of C\,{\sc i} atoms with H\,{\sc i}, H$_2$, and He \cite{Silva2002}. The collision rate coefficients were taken from \cite{Abrahamsson2007} (for H), \cite{Schroder1991} (for ortho- and para-hydrogen), and \cite{Staemmler1991} (for He). The radiative pumping rate coefficients $\Gamma_{\rm01}$ and $\Gamma_{\rm02}$ \cite{Silva2002} were multiplied by the factor $\chi_{\rm UV}$. Since
the C\,{\sc i} column density is low ($\log N\sim14$), the C\,{\sc i} self-shielding effect may be neglected.

\begin{figure*}
\begin{center}
        \includegraphics[width=0.8\textwidth]{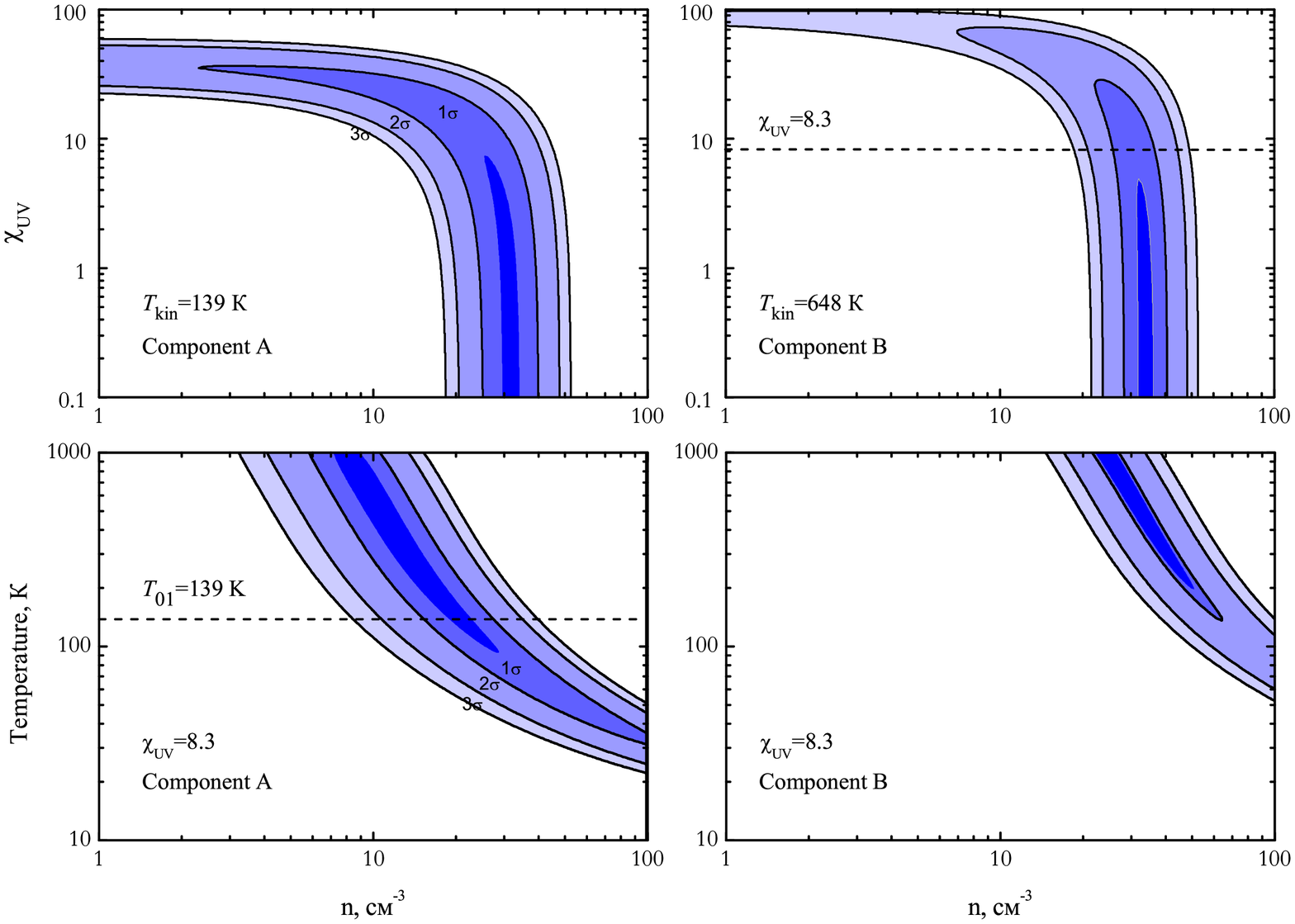}
        \caption{The 1, 2, and 3$\sigma$ confidence regions for determining the physical conditions by analyzing the C\,{\sc i} fine-structure level populations for components A and B (the left and right panels, respectively). The upper panels show the confidence regions for the UV background intensity ($\chi_{\rm UV}$, in units of the mean Galactic value) and the number density. The temperature was assumed to be equal to the value determined from the ortho-to-para-hydrogen ratio, $139$ and $648$\,K for components A and B,
respectively. The lower panels show the confidence regions for the temperature and the number density. The UV background intensity was assumed to be higher than the mean Galactic value by a factor of $8.3$.}
        \label{carbon_diag} 
\end{center}
\end{figure*}

For constant values of the gas molecular fraction and the helium abundance, the relative level populations
${n_{\rm C\,I^*}/n_{\rm C\,I}}$ and ${n_{\rm C\,I^{**}}/n_{\rm C\,I}}$ depend only on the gas number density, temperature, and $\chi_{\rm UV}$. The upper panels in Fig.\,\ref{carbon_diag} show the confidence regions for the number density and the UV background intensity (the gas temperature was assumed to be equal to the values corresponding to the orth-to-para-hydrogen ratio,  $T_{\rm 01}^{\rm A}=139\,K$ and $T_{\rm 01}^{\rm B}=648\,K$, see Section\,\ref{H2inVLT}). The lower panels in Fig.\,\ref{carbon_diag} show the confidence regions for the number density and temperature (the UV background was assumed to be $\chi_{\rm 8.3}$, which corresponds to the background estimate in component B; see the previous section). The helium
abundance for both components was assumed to be $n_{\rm He}/n_{\rm H}=0.083$ (the primordial abundance; see, e.g., \cite{Steigman2007}), while the gas molecular fraction was $f^{\rm A}_{\rm H_2}=0.1$ for component A (the mean value for the sub-DLA system) and $f^{\rm B}_{\rm H_2}=0.001$ for component B.

As shown on the upper panel in Fig.\,\ref{carbon_diag}, the gas number density for component A depends weakly on the UV background if $\chi_{\rm UV}\le 10$. The best value is $n_{\rm A} = 30\pm10$\,cm$^{-3}$. If the UV background in component A is close to the value determined for component B (see the lower panel in Fig.\,\ref{carbon_diag}), then we cannot unambiguously determine the physical conditions: the gas can be hot and rarefied ($T\sim500$\,K and $n\sim10\,\mbox{cm}^{-3}$) or cold and denser ($T\sim50$\,K and $n\sim50\,\mbox{cm}^{-3}$). The accuracy of determining $n$ and $T$ corresponds to a change in the $N_{\rm C\,I^{*}}/N_{\rm C\,I}$ and ${N_{\rm C I^{**}}/N_{\rm C I}}$ ratios within one standard deviation.

For component B, if we assume $T=T_{\rm 01}^{\rm B}\sim650$\,K and $\chi_{\rm UV}\sim8.3$, the gas number density is determined with a high accuracy, $n_{\rm B}\sim26\pm4$\,cm$^{-3}$ (see the upper right panel in Fig.\,\ref{carbon_diag}). However, the ortho-para-hydrogen ratio in the optically thin case is known to be insensitive to the kinetic temperature of the gas (see, e.g., \cite{Abgrall1992}); therefore, the
gas temperature in the cloud can differ from $T_{\rm 01}^{\rm B}$. If the gas temperature is limited, $T\ge100$\,K, then our analysis of the C\,{\sc i} level populations gives an estimate of the gas number density $n\leq100\,\mbox{cm}^{-3}$.

\subsection{The H$_2$ Formation Rate on Dust in Component A}

Under the condition of a stationary balance in the cloud, the numbers of H$_2$ molecules formed on dust
and destroyed by UV radiation are
\begin{equation}
\label{eqH2form}
{\rm R_{H_2}} n_{\rm H} n = 0.11 \beta n_{\rm H_2}, 
\end{equation} 
where ${\rm R_{H_2}}$ is the H$_2$ formation rate coefficient on dust, $n = n_{\rm H} + 2n_{\rm H_2}$ is the total hydrogen number density, and $\beta$ is the photoabsorption rate. The photoabsorption rate in the cloud $\beta$ is related to the intensity of the external UV background by the following relation: $\beta = 4\times10^{-10}\times \chi_{\rm UV}\times S_{\rm shield}(N_{\rm H_2})\,\mbox{\,s}^{-1}$. Usually, the hydrogen molecular fraction in interstellar clouds increases toward the center; therefore, the column density ratio $N^{\rm A}_{\rm H_2}/N_{\rm H\,I} = 0.05$ can be used as a lower limit on $n_{\rm H_2}/n_{\rm H}$ at the cloud center. For the central part of the cloud ($\log N_{\rm H_2} = 17.6$, $S_{\rm shield} = 1.2\times10^{-3}$), we then obtain
\begin{equation}
\label{Rh2eq2}
\frac{{\rm R_{H_2}} n}{\chi_{\rm UV}} =  2.6 \times10^{-15}\times\frac{n_{\rm H_2}/n}{(0.05)}\,\mbox{\,s$^{-1}$} 
\end{equation}  
It follows from our analysis of the C\,{\sc i} level populations that if the UV background in component A
does not exceed $\chi_{\rm UV}\approx10$, then the gas number density is $n_{\rm A} = 30\pm10$\,cm$^{-3}$. We then obtain an estimate of ${\rm R_{H_2}\times\chi_{\rm UV}^{-1} = 8.8\times10^{-17}\,cm^3\,s^{-1}}$. For
comparison, the mean H$_2$ formation rate measured in molecular clouds in our Galaxy is ${\rm R^G_{\rm H_2}= (3-4)\times10^{-17}\,cm^3\,s^{-1}}$ \cite{Jura1975a, Gry2002}. Thus, for the observed amount of H$_2$ to be formed in the sub-DLA system being investigated at the mean Galactic value of ${\rm R_{H_2}}$, the UV background intensity must be a factor of 2.3 lower than the mean Galactic one. However, it follows from our simulations of component A with the Meudon PDR code (see below) that the UV background intensity must be
a factor of $\sim12$ higher than the mean Galactic value. Using this value, we obtained an estimate of the H$_2$ formation rate coefficient on dust in component A, ${\rm R^{A}_{H_2}= 1.1\times10^{-15}\,cm^3\,s^{-1}}$, which is a factor of $\sim35$ higher than the mean Galactic value.

\subsection{The Gas Ionization Fraction}
\label{ion_frac}
For low H$_2$ column densities ($\log N_{\rm H_2}\le20$), absorption in the H$_2$ lines of the Lyman and Werner bands changes insignificantly the number of UV photons capable of destroying HD and ionizing C\,{\sc i} (see, e.g., \cite{Ledoux2015}). Therefore, the HD and C\,{\sc i} abundances in the medium correspond to their equilibrium values, which depend on the e. and H$^+$ number densities, i.e., the gas ionization fraction. 

HD molecules are formed in the reaction of collisions between H$_2$ and D$^+$, while the abundance of D$^+$ ions in a gas with a low molecular fraction is established according to the chemical equilibrium of the
direct and reverse reactions ${\rm H^+ + D \rightleftharpoons H + D^+}$ (see, e.g., \cite{LePetit2002}). Assuming that the cloud is homogeneous and that the HD and H$_2$ molecules belong to the same spatial region, i.e., $N_{\rm HD}/N_{\rm H_2}\approx n_{\rm HD}/n_{\rm H_2}$, we can calculate the number density of H$^+$ ions from the following formula (see, e.g., \cite{LePetit2002}):
\begin{equation}
n_{\rm H^+}=\frac{\beta({\rm HD}) {N_{\rm HD}}}{k_{{\rm H_2 + D^+}} {N_{\rm H_2}} ({\rm D/H}) \frac{k_1}{k_2}},
\end{equation} 
where $\beta({\rm HD})=\chi_{\rm UV}\times1.5\times10^{-11}$\,s$^{-1}$ is the HD photodestruction rate ($\chi_{\rm UV}$ is measured with respect to the intensity of the mean Galactic UV background from \cite{Habing1968}), $k_{\rm H_2 + D^+}=2\times10^{-9}$\,cm$^{3}$s$^{-1}$ is the HD formation rate, ${\rm D/H}=3\times10^{-5}$ is the atomic deuterium abundance for systems at high $z$, and ${k_1/k_2=\exp(-41/T)}$ is the ratio of the direct and inverse ${\rm H^+}$ and  D collision reaction rates. Using the gas temperature estimated from the ortho-to-para hydrogen ratio, $T_{\rm 01} = 140\,$K, and the UV background
intensity estimated from our simulations of component A with the Meudon PDR code, $\chi_{\rm UV} = 12$ (see
Sect.\,\ref{PDRsec}), we obtain $n_{\rm H^+}\simeq0.3$\,cm$^{-3}$.

Neutral carbon is formed in the recombination reaction of C$^+$ ions with electrons and/or polyaromatic
hydrocarbons (PAHs) (see, e.g., \cite{Wolfire2008}). However, the contribution from the recombination
reaction of C$^+$ with PAHs has not been completely established to date (for a discussion and references,
see \cite{Liszt2011}). If this reaction channel is disregarded, then the electron number density in the
gas is expressed in terms of the C\,{\sc i} photodestruction rate $\beta(\rm C)=\chi_{\rm UV}\times2.1\times10^{-10}$\,s$^{-1}$, the C$^+$ recombination coefficient $\alpha({\rm C^+})=1.8\times10^{-11}(T/100)^{-0.83}\mbox{\,cm$^{3}$s$^{-1}$}$ (for $20 K < T < 140 K$;
see \cite{Wolfire2008}), and the ratio of the C\,{\sc i} and C$^+$ number densities:
 \begin{equation}
\label{HD_form}
n_{\rm e^-}=\frac{\beta({\rm C}) {N_{\rm C\,I}}}{\alpha({\rm C^+}){N_{\rm C^+}}},
\end{equation} 
For $\chi_{\rm UV} = 12$ and $T=140\,$K, $\log N_{\rm C\,I} = 14.06$, and $\log N_{\rm C^+} = 16.25$, we obtain $n_{\rm e}\simeq1.2$\,cm$^{-3}$. Assuming that the gas molecular fraction for component
A is $f^{\rm A}_{\rm H_2}= 0.1$ and the number density is $n^{\rm A} = 30$\,cm$^{-3}$, we obtain $n_{\rm e}/n_{\rm H}\sim 4\times10^{-2}$. Thus, in comparison with the values measured in diffuse clouds in our Galaxy with a similar gas number density ($n_{\rm e}/n_{\rm H}\le6.6\times10^{-3}$, see \cite{Welty2003} and $n_{\rm e}/n_{\rm H}\le10^{-6}$), the gas ionization fraction in component A turns out to be higher by almost an order of magnitude.

\section{SIMULATIONS OF THE MOLECULAR CLOUD STRUCTURE}

The estimates of the physical conditions obtained within the homogeneous model (Section\,\ref{sec:physcond}) are often approximate. Fore more proper estimates, we performed simulations of the molecular cloud structure
with the Meudon PDR and CLOUDY codes. 

\subsection{Simulations of the Molecular Clouds with the Meudon PDR Code}
\label{PDRsec}

By comparing the observed H\,{\sc i}, H$_2$, HD, and C\,{\sc i} column densities and the H$_2$ and C\,{\sc i} level population diagrams with the results of our Meudon PDR simulations, we determined the physical conditions in the components of the H$_2$ absorption system. In comparison with the homogeneous model, where the physical conditions and the concentrations of species are assumed to be constant over the cloud volume,
the code takes into account the shielding of UV radiation in absorption lines and the shielding on dust
and consistently solves the thermal, ionization, and chemical balance equations in the medium (by an iterative method). The cloud is represented as planeparallel layers of gas and dust with a constant proton
density in each layer ($n_{\rm p}=n_{\rm H} + 2n_{\rm H_2}+n_{\rm H^+}$).

For the spectrum of the background UV radiation, we used the model proposed by \cite{Mathis1983}. The radiation spectrum spans the wavelength range $912-8000$\,\AA\,. The radiation with a wavelength below
$912$\,\AA\, in the photodissociation region is believed to be shielded by a layer of atomic hydrogen and,
therefore, is disregarded. The UV background intensity with respect to the mean Galactic value
$J^{\rm G}_{\rm UV}\simeq3.2\times10^{-20}\mbox{erg\,s$^{-1}$cm$^2$Hz$^{-1}$rad$^{-1}$}$ taken
from \cite{Habing1968} was specified by the parameter $\chi_{\rm UV}$. In addition to the UV radiation, the cloud is irradiated by the CMBR and the cosmic-ray (CR) background. The CMBR temperature was assumed
to be $2.725\times(1 + z_{\rm abs}) = 8.34$\,K. The CR background intensity with respect to the mean Galactic
value $2\times10^{-16}$\,s$^{-1}$ (see, e.g., \cite{Hollenbach2012}) was specified by the parameter ҐжCR. It is important to note that in the Meudon PDR code the gas is ionized by CRs. Therefore, CR is the key parameter defining the ionization fraction and, consequently, the number of HD molecules and C\,{\sc i} atoms in the cloud being investigated (see Section\,\ref{ion_frac}).

In our simulations, we used the abundances of elements in the cloud (He, C, N, S, Si, Fe, etc.) corresponding to the mean gas metallicity in the sub-DLA system (${\rm [X/H]=-0.2}$; see \cite{Milutinovic2010}). Among these elements, carbon plays a special role in calculating the thermal balance, because the emission from C$^+$ ions in the $\lambda = 158\,\mu{\rm m}$ line is a major gas cooling process in the interstellar medium \cite{Wright1991}. However, since the C\,{\sc ii} absorption line is strongly saturated, we cannot determine what part of the total C$^+$ column density belongs to the molecular clouds being investigated. Therefore, we varied the carbon abundance in the cloud in a range of values corresponding a total carbon column density in the cloud from $\log N^{\rm tot}_{\rm C} = 14.5$ (10$^{-2}$ of the solar abundance) to $\log N^{\rm tot}_{\rm C} = 16.25$ (the upper limit for $\log N^{\rm tot}_{\rm C}$  in the sub-DLA system). At the same time, the abundances of other elements were not varied. The carbon abundance with respect to the solar one, ${\rm (C/H)_{\odot} = 2.7\times10^{-4}}$ \cite{Asplund2009}, was specified by the parameter ${\rm X_C}$. The deuterium abundance was assumed to be ${\rm (D/H) = 3\times10^{-5}}$, which corresponds to the typical D abundance at high redshifts in molecular clouds
(see, e.g., \cite{Ivanchik2015}).

Dust is one of the most important components in the interstellar medium; its properties affect the heating
rate of the medium (through the photo-electric reaction on dust), the gas shielding from UV radiation,
and the H$_2$ formation rate. The model from \cite{Mathis1977} was used for the grain size distribution.
The absorption of UV radiation on dust was described using an extinction model for the SMC, $ R_{\rm V} = 2.87$ (this model is believed to describe well the reddening of quasar spectra; see, e.g., \cite{Noterdaeme2009b}). The amount of dust on the line of sight was determined via the interstellar
extinction parameter $A_{\rm V} = R_{\rm V}E({\rm B-V})$, which was set equal to $0.115\,mag$, corresponding to a certain upper limit on the spectrum reddening (color excess), $E({\rm B-V}) = 0.04$ \cite{Ledoux2015}. The dust number density in the cloud governing the gas heating and the H$_2$ formation rate is determined using the dust-to-gas ratio $G$. We used three values: $G =0.01$, $0.02$, and $0.1$, corresponding to 1, 2, and 10 mean Galactic values \cite{Bohlin1978}. The H$_2$ formation in the Meudon PDR code is computed using two mechanisms: Langmuir–Hinshelwood and Eley–Rideal (see \cite{LeBourlot2012}). However, as is pointed out in \cite{LeBourlot2012}, the detailed description of H$_2$ formation is still missing. Therefore, we used two approaches: (i) a ``numerical calculation'' according to the Langmuir–Hinshelwood and Eley–Rideal models and (ii) an ``approximate calculation'' where the H$_2$ formation rate was specified by some value constant over the cloud volume:
${\rm R_{H_2}=R_0\times3\times10^{-17}\mbox{\,cm$^3$s$^{-1}$}}$
(with respect to the mean Galactic value; see \cite{Jura1975a}). The latter approach allows the possibility of a difference between the properties of dust in a high-redshift cloud and in our Galaxy to be taken into account qualitatively without going into the specific properties of dust grains (the size distribution, the reflection coefficient, the sticking coefficient, the density, etc.).'

\subsubsection{Simulations of the cloud J\,2123$-$0050\,A}

The DLA systems have much larger sizes than the molecular clouds: the sizes of the DLA systems
are estimated to be several 10\,kpc \cite{Krogager2012}, while the sizes of the H$_2$ clouds are $\sim1$\,pc (see, e.g., \cite{Noterdaeme2007}). Therefore, the DLA system occupies a larger region than does the H$_2$ system on the line of sight (with the H$_2$ cloud being located inside the DLA system). Consequently, we cannot reliably determine what fraction of the entire HI column density belongs to the H$_2$ cloud by measuring the total HI and H2 column densities. Since the sub-DLA system being investigated has the lowest H\,{\sc i} column density (almost two orders of magnitude lower than the mean $N_{\rm H\,I}$; see Fig.\,\ref{H2vsHI}) among the sub-DLA and DLA systems containing H$_2$ and, at the same time, the H$_2$ column density in component A is fairly high (close to the mean $N_{\rm H_2}$), in our simulations we assume that the entire observed H\,{\sc i} in the sub-DLA system belongs to the H$_2$ cloud. Therefore, the total hydrogen column density ($N_{\rm H\,I}$ + 2$N_{\rm H_2}$)  was taken to be $\log N=19.22$.

We used two thermodynamic cloud models: (I) with a constant pressure and (II) with a constant hydrogen density. The pressure was assumed to be $(8.4\times10^3)\mbox{\,cm$^{-3}$\,K}$, which corresponds to the upper limit determined by analyzing the C\,{\sc i} level populations. The hydrogen number density was assumed
to be 40\,cm$^{-3}$, which roughly corresponds to the conditions in the cloud. The second model describes
the situation where H$_2$ is formed in the compression region, which was formed, for example, through the
gas interaction with the shock front.

\begin{figure*}
\begin{center}
        \includegraphics[width=1.0\textwidth]{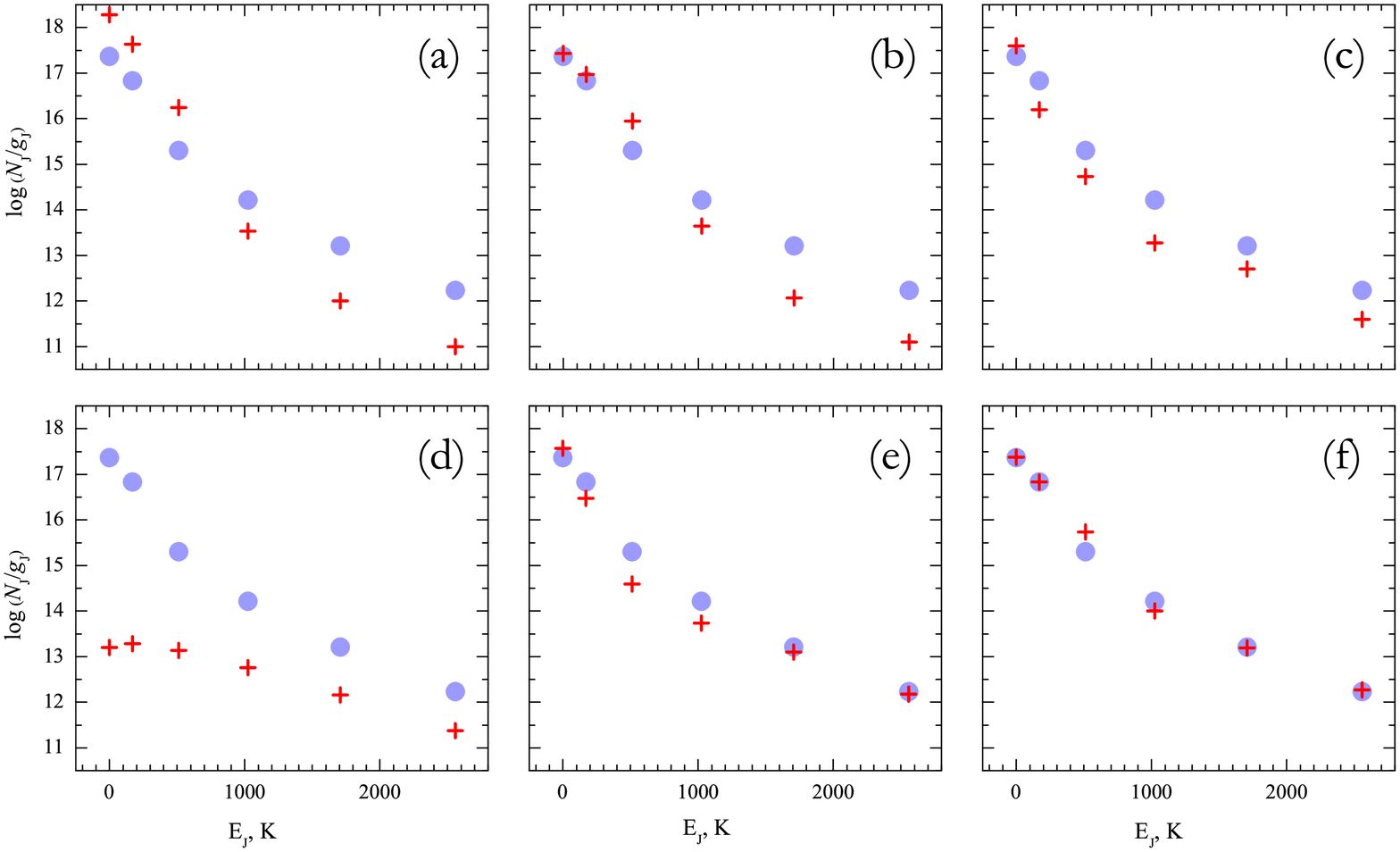}
        \caption{Comparison of the measured H$_2$ level populations for component A (gray circles) with the results of our PDR Meudon simulations (red crosses) for six models (a)–--(f), whose parameters are given in Table\,\ref{PDRtable}.}
        \label{PDRgraph1} 
\end{center}
\end{figure*}

For each model, we performed our calculations for a grid of parameters $\chi_{\rm UV}$, $\zeta_{\rm CR}$, ${\rm R_{0}}$, $X_{\rm C}$ and $G$. The model with a constant density gives the best description for the following set of parameters:  $\chi_{\rm UV} =12$, $\zeta_{\rm CR}=5\times10^2$, ${\rm R_{0}=38}$, $X_{\rm C}=2.8$ and $G=0.01$ (model (f) in Table\,\ref{PDRtable}). The model simultaneously reproduces the observed column densities of the species and the H$_2$ and C\,{\sc i} level populations. Table\,\ref{PDRtable} and
Fig.\,\ref{PDRgraph1} compare the results of our simulations for the six most characteristic cases: (a--e) correspond to the isobaric models, (f) corresponds to the model with a constant density.

\begin{table*}
\begin{center}
\caption{Results of our simulations of the absorption system J 2123-0050 A. $\chi_{\rm UV}$, $\zeta_{\rm CR}$ and ${\rm R_0}$ --  are the UV background intensity, the CR background intensity, and the H$_2$ formation rate with respect to their mean Galactic values; $X_{\rm C}$ is the carbon abundance with respect to the solar one; $n$ and $T$ are the number density and temperature at the
cloud center. $^*$\,--\,The range of the formation rate coefficient ${\rm R_0}$ calculated numerically according to the model from \cite{LeBourlot2012} is specified.} 
\label{PDRtable}
\begin{tabular}{|c|c|c|c|c|c|c|c|} 
\hline 
No.& $\chi_{\rm UV}$ & $\zeta_{\rm CR}$ & $X_{\rm C}$ & ${\rm R_{0}}$  & $n$, cm$^{-3}$ & $T$, K & $N_{\rm H^+}/N_{\rm H}$\\
\hline                                   
a & 0.4 & 1 & 0.02 & 1.7-3.0$^*$  &  73 & 115 & $4.7\times10^{-4}$\\ 
b & 1.2 & 1 & 0.2  &  1.7-2.2$^*$ & 47  & 180 & $6.0\times10^{-3}$ \\ 
c & 5  & 1 & 1.8 &  1.7-1.9$^*$ & 165 & 51 & $1.5\times10^{-4}$ \\ 
d & 12 & 5 & 3.7 & 0.5-0.9$^*$ & 180 & 47 & $1.7\times10^{-4}$  \\  
e & 12 & $5\times10^2$ & 3.0 &  10 & 124 & 68 & $4.7\times10^{-3}$  \\ 
\hline
f & 12 & $5\times10^2$ & 2.8 &  38 & 40 & 145 & $1.1\times10^{-2}$\\ 
\hline
\hline
No. & $\log N_{\rm H_2}$ & $\log N_{\rm C\,I}$ & $\log N_{\rm HD}$ & $N_{\rm H_2}^{\rm J=1}/N_{\rm H_2}^{\rm J=0}$ &$\log N_{\rm H_2}^{\rm J=4}$ & $\log N_{\rm H_2}^{\rm J=5}$ & $N_{\rm C\,I^*}/N_{\rm C\,I}$\\
\hline
a & 18.8 & 11.4 & 14.2 & 2.0 & 13.0 & 12.5 & 0.4 \\ 
b & 18.0 & 11.7 & 13.5 & 3.1 & 13.0 & 12.6 & 0.4  \\
c & 17.7 & 13.7 & 12.8 & 0.4 & 13.7 & 13.1 & 0.7  \\
d & 14.6 & 14.0 & 10.6 & 1.1 & 13.1 & 12.9 & 0.7   \\
e & 17.8 & 14.3 & 13.9 & 0.7 & 14.1 & 13.7 & 0.7  \\
\hline
f & 17.9 & 14.0 & 14.0 & 2.6 &  14.1 & 13.8 & 0.4 \\
\hline
\end{tabular}
\end{center}
\end{table*}   

\begin{table*}
\begin{center}
\caption{Results of our simulations of the absorption system J\,2123$-$0050\,B. The notation is the same as that in Table\,\ref{PDRtable}. $^*$\,--\,The range of the formation rate coefficient ${\rm R_0}$ calculated numerically according to the model from \cite{LeBourlot2012} is specified.} 
\label{PDRtable2}
\begin{tabular}{|c|c|c|c|c|c|c|c|c|c|} 
\hline
No.  & $\chi_{\rm UV}$ & $\zeta_{\rm CR}$ & ${\rm R_{0}}$ & $X_{\rm C}$ & $\log N_{\rm H}^{\rm tot}$ & $\log N_{\rm H_2}$ & $\log N_{\rm C\,I}$  & $\log N_{\rm H_2}^{\rm J=4}$ & $\log N_{\rm H_2}^{\rm J=5}$\\  
\hline
g & 1.5 & 1 & { 5.2-8.9}$^*$ & $0.02$ & 19.2 & 15.7 & 9.8 & 12.7 & 12.3\\
h & 12 & 1 & 43 & $2.8$ & 18.7 & 15.2 & 12.0 & 13.6 & 13.4\\ 
i & 12 & $1.5\times10^2$ & 43 & $2.8$ & 18.7 & 15.2 & 12.9 & 13.6 & 13.4\\ 
\hline
\end{tabular}
\end{center}
\end{table*}   

\begin{figure*}
\begin{center}
        \includegraphics[width=1.0\textwidth]{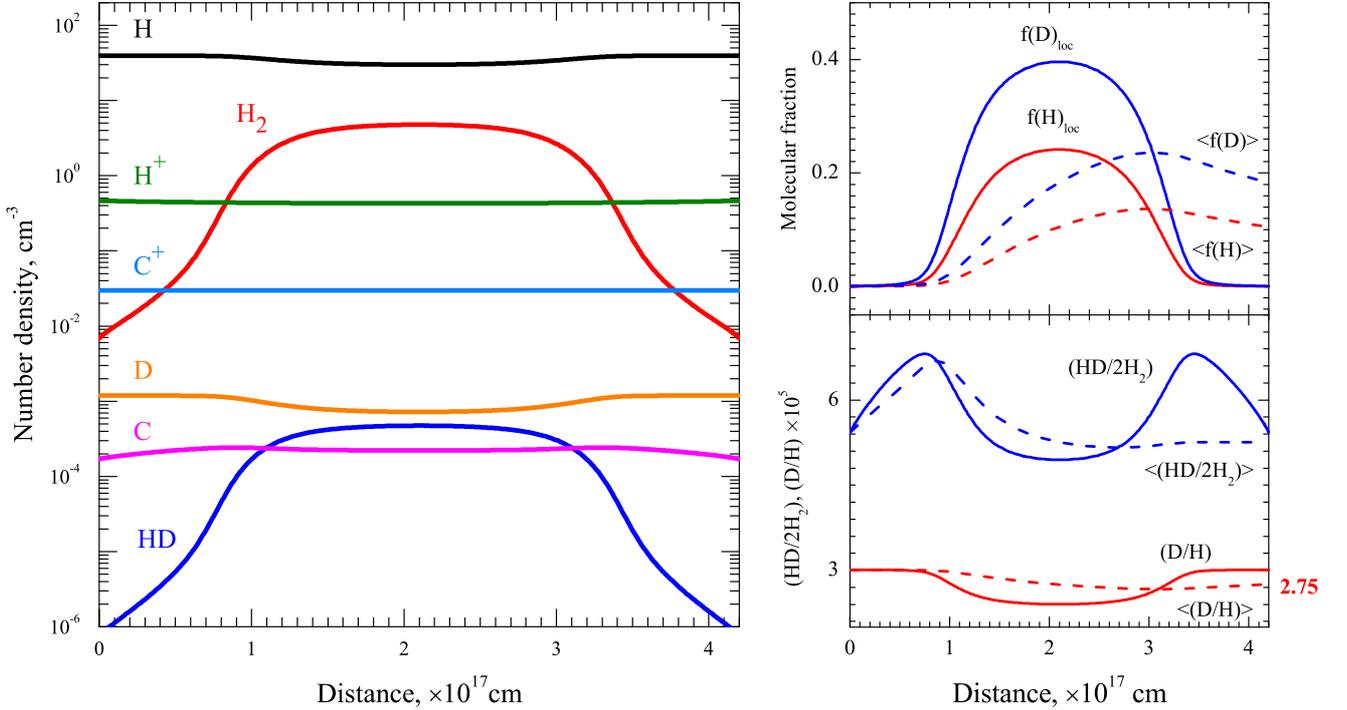}
        \caption{Structure of the molecular cloud obtained in the PDR Meudon model (model (f), see Table\,\ref{PDRtable}). The left panel shows the number density profiles for the species in the cloud. The solid and dashed lines on the right panels indicate the dependences of the absolute and mean H and D molecular fractions (upper panel) as well as the $N_{\rm HD}/2N_{\rm H_2}$ and $N_{\rm D}/N_{\rm H}$ ratios (lower panel). The averaging was performed from one of the cloud boundaries.}
        \label{PDRgraph2} 
\end{center}
\end{figure*}

\begin{figure*}
\begin{center}
        \includegraphics[width=1.0\textwidth]{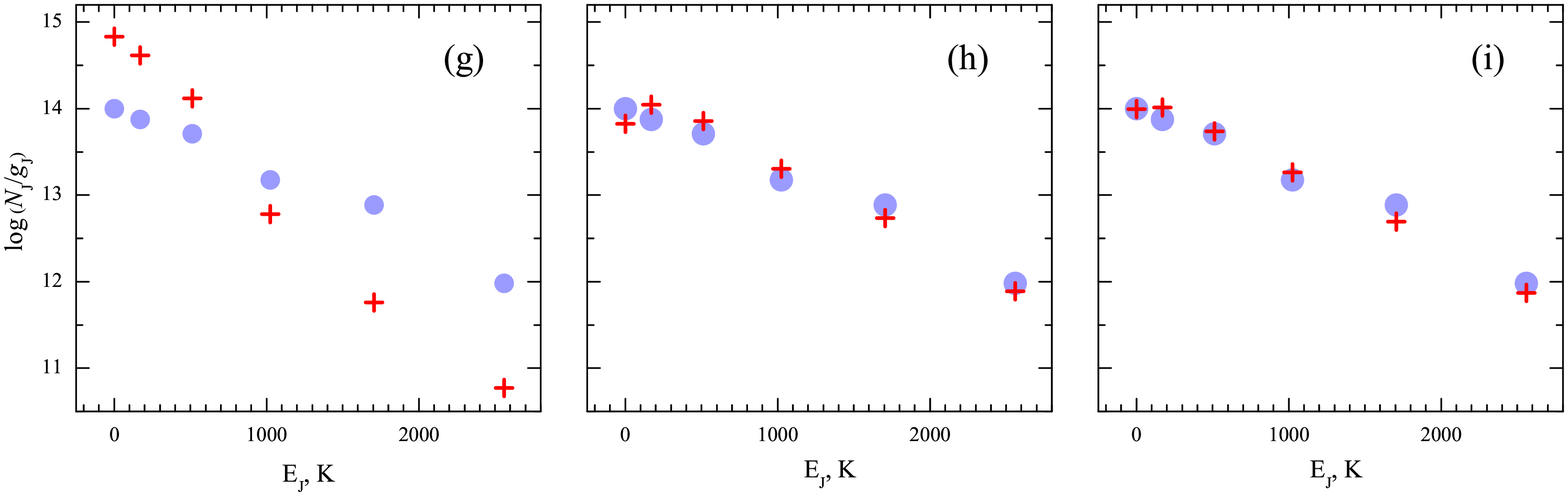}
        \caption{Comparison of the measured H$_2$ level populations for component B (gray circles) with the results of our PDR Meudon simulations (red crosses) for three models (g)--(i), whose parameters are given in Table\,\ref{PDRtable2}.}
        \label{PDRgraph3} 
\end{center}
\end{figure*}

{\bf Main inconsistencies of the isobaric models.}

Using an isobaric model, we can obtain H$_2$, HD, and C\,{\sc i} column densities close to the observed ones (models (с) and (e)), but we fail to satisfactorily describe the H$_2$ and C\,{\sc i} level population diagrams. 

Models (a) and (b) correspond to the case where the medium is irradiated by an UV background with
a low intensity, $\chi_{\rm UV} = 0.4$ and 1.2. In these models, the populations of H$_2$ levels with J$\ge$3 are more than an order of magnitude lower than the observed values, despite the fact that the total H$_2$ column density is close to or even higher than the measured value. An additional level population can be provided through radiative pumping with a higher UV background intensity\footnote{Apart from the population mechanism for high H$_2$ levels through radiative pumping, other mechanisms that are disregarded
in the Meudon PDR code are considered in the literature. For example, the excitation of H$_2$ molecules when
the gas is compressed and heated due to the interaction of the molecular cloud with turbulence and/or a C--shock (see \cite{Joulain1998, CecchiPestellini2005, LePetit2004}).}\,. Note that the column density of
HD molecules can be described in these models, but the C\,{\sc i} column density turns out to be lower by more than two orders of magnitude.

Model (с) describes a cloud irradiated by an UV background with an intensity higher than the mean Galactic one by a factor of 5. A high UV background increases the gas heating, which in this model is compensated for by an enhanced carbon abundance (${\rm X_C = 1.9}$), because carbon is the main coolant of the medium. The model yields H$_2$ and C\,{\sc i} column densities close to the observed ones. However, the derived HD column density and the H$_2$ and C\,{\sc i} level populations are inconsistent with the measured values.
The populations of the J = 4 and 5 levels turn out to be a factor of 3 higher than those in models (a) and
(b), but still a factor of $2-3$ lower than the observed values.

In model (d), the radiation intensity is even higher: $\chi_{\rm UV}\simeq12$. To compensate for the cloud heating, the carbon abundance was taken to be maximally high, ${\rm X_C = 3.7}$ (which corresponds to the case where the entire C$^+$ column density in the sub-DLA system belongs to the molecular cloud). The gas number density and temperature be found the same as those in model (c). However, the obtained column density of H$_2$ in model (d) is $\log N_{\rm H_2}\sim15$, that is less than the measured value by two orders of magnitude. First, the H$_2$ column density is too low to shield the molecules from UV radiation; second, at the same H$_2$ formation rate as that in model (c), the photodestruction rate turns out to be higher (see Eq.\,(\ref{eqH2form})).

In models (a)--(d), we used the ``numerical calculation'' according to the model from \cite{LeBourlot2012}, with the dust-to-gas ratio in the medium $G = 0.01$, for the H$_2$ formation. As $G$ increases, the H$_2$ formation rate coefficient and the efficiency of gas heating by UV radiation increase proportionally. Despite the fact that the H$_2$ formation rate becomes higher, the gas number density decreases (due to the pressure constancy condition). Therefore, according to Eq.\,(\ref{eqH2form}), increasing the dust content does not allow the problem of the formation of a large amount of molecular hydrogen in a medium with a high UV background to be solved. In our calculations using an isobaric model with a high UV background, a high carbon abundance (as in model (d)), and $G = 0.02$
and $0.1$, the gas turns out to be hot, $T>1000$\,K, and rarefied, $n\le1$\,cm$^{-3}$. Virtually no H$_2$ molecules are formed in such a medium.

The ``approximate calculation'', where the H$_2$ formation rate coefficient is specified by a value constant
over the cloud volume and does not depend on the physical conditions, can be used for the H$_2$ formation.
Model (e) corresponds to this case. At ${\rm R_{0}\simeq10}$, $\chi_{\rm UV}=12$ and $X_{\rm C}=3$  we can simultaneously reproduce the total H$_2$ column density and the upper level populations (see Table\,\ref{PDRtable}). However, in this model we fail to reproduce the physical conditions in the cloud: the gas turns out to be denser ($n\ge100$\,cm$^{-3}$) and colder ($T\le60$\,K). As a consequence, the ratio of the populations of the first two H$_2$ and C\,{\sc i} levels is consistent with the measured H$_2$ number
densities, while the ratio of the C\,{\sc i}$^*$ and C\,{\sc i} level populations is twice that measured in the component A.

{\bf The abundance of HD molecules and the gas ionization fraction}

For models (a)--(d), the HD/2H$_2$ ratio does not exceed $10^{-5}$, which is less than the observed value by almost an order of magnitude. Since HD is formed through the ion--molecular reaction (${\rm H_2 + D^+}$), the HD formation rate increases with H$^+$ number density, which in the PDR models is determined mainly by the CR background intensity. The HD/2H$_2$ ratio is consistent with the measured value at a relative H$^+$ number density $n_{\rm H^+}/n_{\rm H}\approx10^{-2}$, which corresponds to $\zeta_{\rm CR} = 5\times10^2$, i.e., a factor of 500 more intense CR flux is needed. For $\zeta_{\rm CR} = 1$, the gas ionization fraction in the Meudon PDR models is low, $n_{\rm H^+}/n_{\rm H}\approx10^{-4}$, which is not enough for the HD formation in a medium with a high UV background intensity. A high $\zeta_{\rm CR}$ is probably not an estimate of the CR background intensity in the cloud, but only points to a high gas ionization fraction.
Possible causes are discussed below in Section\,\ref{cloudy}.

{\bf The model with a constant density}

In the model with a constant density (model (f)), we can simultaneously describe the H$_2$, HD, and C\,{\sc i} column densities and the H$_2$ and C\,{\sc i} level populations. In this case, just as for the isobaric model, high UV and CR background intensities and a high carbon abundance are needed.

 A detailed analysis of model (f) is presented in Fig.\,\ref{PDRgraph2}. The left panel shows the number densities of the species as a function of the distance from the cloud boundary. The H, H$^+$, C$^+$, and C\,{\sc i} number densities barely change along the line of sight (due to the density constancy and the low gas molecular fraction). The right panels show the D and H molecular fractions (the upper panel: the local values are indicated by the solid lines, while the values averaged along the line of sight are indicated by the dashed lines) and the HD/2H$_2$ and D/H ratios (the lower panel). The HD formation rate in the cloud is higher than the H$_2$ formation rate due to the relatively high gas ionization fraction, $\sim10^{-2}$. As a result, the local D and H molecular fractions at the cloud center turn out to be twice their mean values, and, consequently, the HD/2H$_2$ ratio is higher than the D/H ratio. This leads to the D/H abundance determined from the ratio of the atomic D\,{\sc i} and H\,{\sc i} column densities being lower than the isotropic D/H ratio ($2.75\times10^{-5}$ versus $3.0\times10^{-5}$). The sign and magnitude of the effect depend on the mean gas molecular fraction along the line of sight and the H$^+$ number density. For $f_P{\rm H_2} = 0.1$, the $N_{\rm D\,I}/N_{\rm H\,I}$ ratio can change within 10\,\% of the isotopic ratio. The effect should be taken into account if a molecular cloud with the same redshift falls on the line of sight passing through the H\,{\sc i}/D\,{\sc i} absorption system. Obviously, in this case, the $(N_{\rm D\,I}+N_{\rm HD})/(N_{\rm H\,I}+2N_{\rm H_2})$ ratio is a proper estimate of (D/H).
 
\subsubsection{Simulations of the cloud J\,2123$-$0050\,B}

For component B (just as for component A), we cannot determine what part of the measure H\,{\sc i}
column density associated with the molecular cloud. Therefore, $N^{\rm B}_{\rm H\,I}$ was an additional free parameter and was determined from the simulation results. The H\,{\sc i} column density in the sub-DLA system, $\log N_{\rm H\,I} = 19.18\pm0.15$, was chosen as an upper limit for $N^{\rm B}_{\rm H\,I}$. The results of our simulations are presented in Table\,\ref{PDRtable2}. Model (g) corresponds to the calculations
with a constant gas pressure ($8.4\times10^3$\,cm$^{-3}$\,K); models (h) and (i) correspond to those with a constant number density (40\,cm$^{-3}$). Figure\,\ref{PDRgraph3} shows the H$_2$ level population diagrams for these models. Just as in the case of simulating component A, a high UV background intensity ($\chi_{\rm UV}\sim12$) is needed to describe the populations of upper rotational H$_2$ levels. In model (g), the UV background is low and the H$_2$ levels are populated mainly through the collisional mechanism. The populations of the lower H$_2$ levels turn out to be a factor of $3-5$ higher than the observed ones, while the populations of the upper J = 4 and 5 levels are an order of magnitude lower. If the UV background is high (models (h) and (i)), then the H$_2$ level population diagram describes the diagram measured for component B.

At a high UV background and a low gas number density, a high H$_2$ formation rate coefficient on dust
is needed for the H$_2$ formation. In model (g), the H$_2$ formation was calculated according to \cite{LeBourlot2012}; in models (h) and (i), the formation rate coefficient was equal to ${\rm R_0}$. For ${\rm R_0 = 43}$, the H$_2$ column density in models (h) and (i) turns out to be close to the observed one. The gas ionization fraction in component B can be estimated from the C\,{\sc i} abundance (see, e.g., Section\,\ref{ion_frac}). The observed C\,{\sc i} column density in model (i) is reconstructed at $n_{\rm e}/ n_{\rm H}=(0.9-1)\times10^{-2}$, which corresponds to $\zeta_{\rm CR}=1.5\times10^2$.

\subsection{The Ionization Structure of the sub-DLA System. The CLOUDY Code}
\label{cloudy}

The high value of $\zeta_{\rm CR}$ ($5\times10^2$ for component A and $1.5\times10^2$ for component B) obtained in our simulations with the Meudon PDR code seems unlikely and probably points only to a high gas ionization that can actually be caused by other sources, for example, hard UV or X-ray radiation, which cannot be taken into account in the simulations with the Meudon PDR code. However, this can be done
with the CLOUDY code, which can compute the ionization of the system by taking into account the
hard UV radiation. We simulated the structure of the sub-DLA system and checked whether the external
UV radiation (which is shielded incompletely in the system due to the low H\,{\sc i} column density) could cause a high gas ionization fraction ($\sim10^{-2}$) at the system's center, where the molecular cloud could be located.

\begin{figure*}
\begin{center}
        \includegraphics[width=0.8\textwidth]{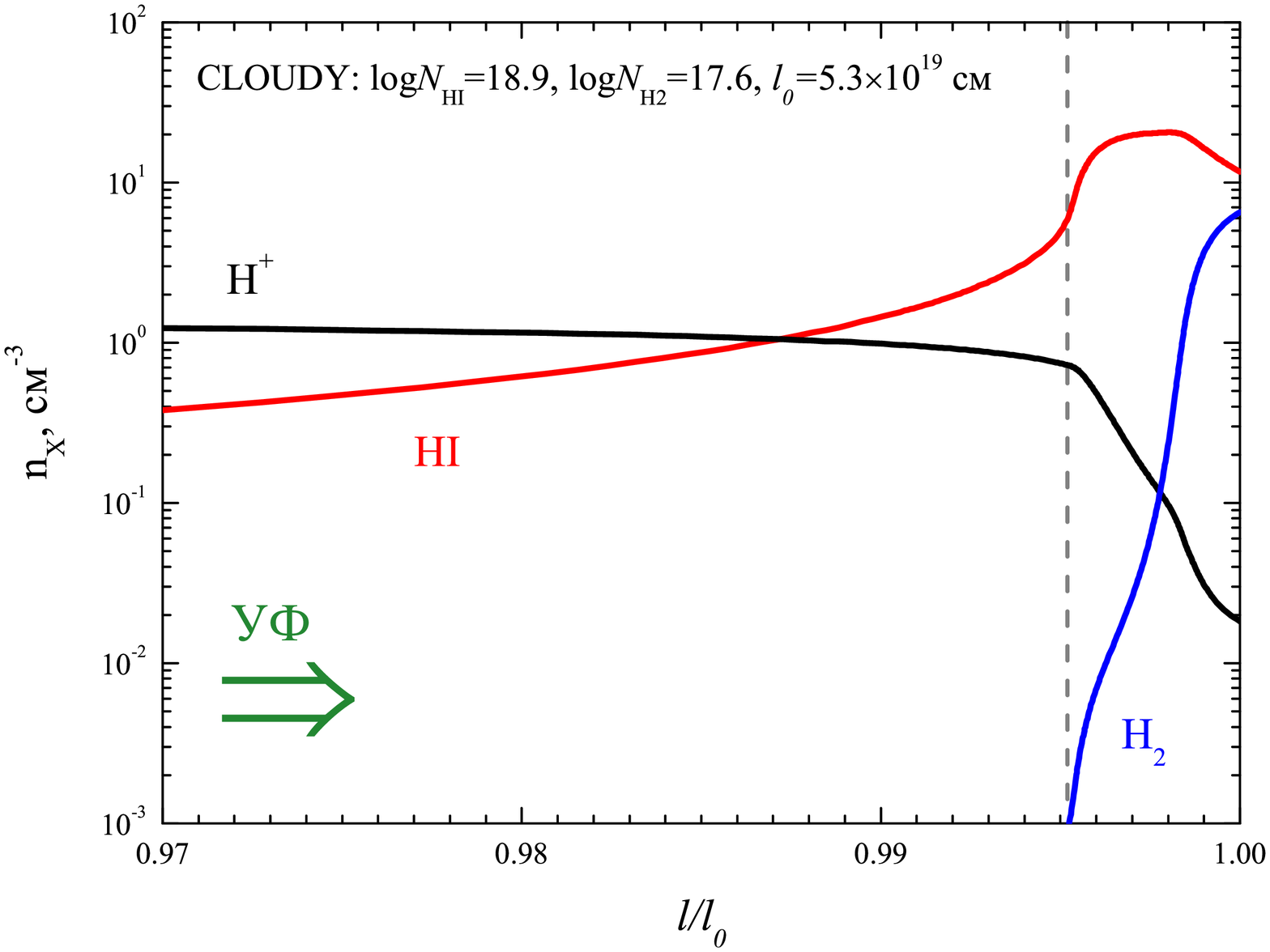}
        \caption{Results of our simulations of the structure of the sub-DLA system with the CLOUDY code. The dependences of the H$^+$, H\,{\sc i}, and H$_2$ number densities on the distance from the boundary are shown. The calculations were performed in a plane-parallel model. The medium is irradiated by UV radiation only on the left side. The vertical line indicates an arbitrary boundary between the regions of hot ionized and cold neutral media.}
        \label{cloudy_sim} 
\end{center}
\end{figure*}

 The computation was performed within the framework of an isobaric model with a plane--parallel
geometry. The gas is irradiated only on one side of the system. The radiation spectrum consists of several
components: (i) the extragalactic background (the HM96 model; see \cite{HM1996}), (ii) the average Galactic background (the ``ISM'' model) with an intensity that is a factor of 12 higher than $J^{\rm G}_{\rm UV}\simeq3.2\times10^{-20}\mbox{erg\,s$^{-1}$cm$^2$Hz$^{-1}$rad$^{-1}$}$ \cite{Habing1968},
and (iii) the CMBR at $z = 2$. The CR background intensity was equal to the mean Galactic value $2\times10^{-16}\mbox{\,с}^{-1}$ (see, e.g., \cite{Hollenbach2012, Liszt2015}). The abundances of
heavy elements corresponded to themetallicity for the sub-DLA system,  ${\rm [S/H]=-0.2}$. Our simulations
with the CLOUDY code also show the necessity of a high H$_2$ formation rate, which was assumed to be $2.3\times10^{-15}\,{\rm cm^3\,s^{-1}}$. Since the total hydrogen column density $N_{\rm H^+}+N_{\rm H\,I}+2N_{\rm H_2}$ in the sub-DLA system cannot be determined (because H$^+$, which accounts for as much as 90\,\%, is undetectable), the computation was performed until the total neutral hydrogen column density $N_{\rm H\,I}+2N_{\rm H_2}$ reached $10^{18.9}$ (half of $N^{\rm tot}_{\rm H\,I}$ for the sub-DLA being investigated). The result of our simulations is shown in Fig.\,\ref{cloudy_sim}. The H$^+$, H,{\sc i}, and H$_2$ number density profiles are shown as a function of the distance from the system’s center.

The total H$^+$ column density is $\log N_{\rm H^+} = 19.9$, which is a factor of 10 higher than the H\,{\sc i} column density, i.e., the neutral gas is formed only in a narrow region near the system’s center, while the bulk of the system ($\sim90\,\%$) is ionized. The number density of H$^+$ ions decreases rapidly with distance from the boundary between the cold neutral and hot ionized phases (see Fig.\,\ref{cloudy_sim}) and changes from $2.0$\,cm$^{-2}$ at the boundary to $2\times10^{-2}$\,cm$^{-2}$ at the system’s center. The medium is ionized through its irradiation by the UV background; the CR contribution is negligible
(the H$^+$, H\,{\sc i} and H$_2$ profiles barely change if the CR background is disregarded). Thus, despite the fact that the ion number density decreases rapidly with neutral cloud depth, the ionization fraction of the medium remains fairly high because of the small cloud size. The mean $n_{\rm H^+}/n_{\rm H}\sim10^{-2}$, in agreement with the value derived in our Meudon PDR simulations for components A and B. Thus, partial shielding of the hard UV radiation by neutral hydrogen in the sub-DLA system can be responsible for the high
gas ionization fraction found in our Meudon PDR simulations.

\section{CONCLUSION}
We performed an independent analysis of the H$_2$ absorption system at $z_{\rm abs} = 2.059$ in the VLT
spectrum of the quasar J\,2123$-$0050. The H$_2$ system consists of two components ($z_{\rm A} = 2.05932(5)$
and $z_{\rm B} = 2.05955(2)$) with total column densities $\log N_{\rm H_2}^{\rm A}=17.94\pm0.01$ и $\log N_{\rm H_2}^{B}=15.16\pm0.01$. The lines of HD molecules are detected only in component A with a column density $\log N_{\rm HD}^{\rm A}=13.87\pm0.06$. Our estimate of the H$_2$ column density in component A is twice the value, obtained by \cite{Tumlinson2010}, $\log N_{\rm H_2}^{\rm A}=17.64\pm0.15$. The results
differ mainly by the estimate of the H$_2$ column density in component A at the J = 0 level. Using our
estimate of $\log N^{\rm A}_{\rm H_2}$, we calculated the abundance of HD molecules: $N^{\rm A}_{\rm HD}/2N^{\rm A}_{\rm H_2}=(4.26\pm0.60)\times10^{-5}$. This value is consistent, within the statistical error
limits, with that measured by \cite{Tumlinson2010}, but it exceeds the primordial deuterium abundance
${\rm (D/H)_{Pr}}=(2.62\pm0.15)\times10^{-5}$ estimated by \cite{Planck2015} by almost three standard deviations, which may be due to the complex chemistry of the H$_2$ and HD molecules.

For component A, we detected the partial coverage effect for the H$_2$ lines located near the Ly$\beta$ and O\,{\sc vi} emission lines. The residual flux is $\sim3$\,\% of the total flux. Due to the low H$_2$ column density and low oscillator strength for transitions of L1-0 and L0-0 Layman bands, allowance for the partial coverage effect affects weakly the H$_2$ column density being determined.

The C\,{\sc i} lines associated with the H$_2$ system for the transitions from the ground and two excited states are detected in the VLT spectrum. Our analysis showed that the C\,{\sc i} line structure consists of three components, each of which is associated with components A and B. The third additional component (with
$\log N^{\rm C}_{\rm C\,I}=12.78\pm0.03$ and a large Doppler parameter $b\sim7$\,km/s) probably associated with the ionized part of the sub-DLA system and not with the H$_2$ system.

Using the observed H\,{\sc i}, H$_2$, HD, and C\,{\sc i} column densities as well as the H$_2$ and C\,{\sc i} level population diagrams for components A and B, we investigated the physical conditions in this system. The medium in component A is optically thick in H$_2$ lines and has a low number density ($30\pm10\,$cm$^{-3}$) and temperature ($T_{\rm 01}=139\pm6$\,K). In such a medium, molecular hydrogen shields the UV radiation in lines in the cloud (the H$_2$ photodestruction rate inside and at the
edge of the cloud can differ by several orders of magnitude); therefore, we used our simulations
with the Meudon PDR code to properly determine the physical conditions with allowance made for the
UV background transfer inside the cloud. Using the model with a constant gas density $n = 40$\,cm$^{-3}$,
a high UV background ($\chi_{\rm UV}\sim12$ in units of the mean Galactic value), a high gas ionization fraction ($n_{\rm H^+}/n_{\rm H}\sim10^{-2}$), and a high H$_2$ formation rate coefficient
on dust ($\rm R_{H_2}=1.2\times10^{-15}$\,cm$^{3}$s$^{-1}$), we can describe the observed parameters.

The high value of $N_{\rm HD}/2N_{\rm H_2}$ in component A is due to the high gas ionization fraction ($\sim10^{-2}$), which is probably a unique feature of this sub-DLA system. Our simulations with the CLOUDY code show that such an ionization fraction can be a consequence of incomplete shielding of the UV radiation (with ${\rm E > 13.6}$\,eV) by neutral hydrogen in the sub-DLA system. A high gas ionization fraction leads to a significant increase in the HD formation rate. Therefore, even at a low gas molecular fraction, $f_{\rm H_2}\sim0.1-0.2$, the deuterium molecular fraction turns out to be higher than the hydrogen molecular fraction, which leads to the relation $N_{\rm HD}/2N_{\rm H_2}>{\rm (D/H)}$.

The medium in component B is optically thin in H$_2$ lines. In this case, the H$_2$ photodestruction rate
barely changes with the depth of radiation penetration into the cloud. Using the homogeneous model,
we estimated the UV background intensity in the cloud to be $\chi_{\rm UV}=8.3$. Our simulations with the
Meudon PDR code for this component showed that the same high H$_2$ formation rate coefficient on dust as
that in component A ($\rm R_{H_2}=1.3\times10^{-15}$\,cm$^{3}$s$^{-1}$) and a high gas ionization fraction ($\sim10^{-2}$) are needed to produce the observed amount of H$_2$ and C\,{\sc i} at a high
UV background ($\chi_{\rm UV}=12$).

\section{ACKNOWLEDGMENTS}
\label{ACKNOWLEDGMENTS}
\noindent
This study was  supported by the Russian Science Foundation (project no. 14-12-00955).


\bibliographystyle{splncs}
\bibliography{SK}

\end{document}